# Gamma-ray Light Curves and Variability of Bright Fermi-Detected Blazars

A. A. Abdo[1,2], M. Ackermann[3], M. Ajello[3], E. Antolini[4,5], L. Baldini[6], J. Ballet[7], G. Barbiellini[8,9], D. Bastieri[10,11], K. Bechtol[3], R. Bellazzini[6], B. Berenji[3], R. D. Blandford[3], E. D. Bloom[3], E. Bonamente[4,5], A. W. Borgland[3], A. Bouvier[3], J. Bregeon[6], A. Brez[6], M. Brigida[12,13], P. Bruel[14], R. Buehler[3], T. H. Burnett[15], S. Buson[10], G. A. Caliandro[16], R. A. Cameron[3], P. A. Caraveo[17], S. Carrigan[11], J. M. Casandjian[7], E. Cavazzuti[18], C. Cecchi[4,5], Ö. Çelik[19,20,21], A. Chekhtman[1,22], C. C. Cheung[1,2], J. Chiang[3], S. Ciprini[5*], R. Claus[3], J. Cohen-Tanugi[24], L. R. Cominsky[25], J. Conrad[26,27,28], L. Costamante[3], S. Cutini[18*], C. D. Dermer[1], A. de Angelis[29], F. de Palma[12,13], E. do Couto e Silva[3], P. S. Drell[3], R. Dubois[3], D. Dumora[30,31], C. Farnier[24], C. Favuzzi[12,13], S. J. Fegan[14], W. B. Focke[3], P. Fortin[14], M. Frailis[29,32], Y. Fukazawa[33], S. Funk[3], P. Fusco[12,13], F. Gargano[13], D. Gasparrini[18], N. Gehrels[19], S. Germani[4,5], B. Giebels[14], N. Giglietto[12,13], P. Giommi[18], F. Giordano[12,13], T. Glanzman[3], G. Godfrey[3], I. A. Grenier[7], M.-H. Grondin[30,31], J. E. Grove[1], S. Guiriec[34], D. Hadasch[35], M. Hayashida[3], E. Hays[19], S. E. Healey[3], D. Horan[14], R. E. Hughes[36], R. Itoh[33], G. Jóhannesson[3], A. S. Johnson[3], W. N. Johnson[1], T. Kamae[3], H. Katagiri[33], J. Kataoka[37], N. Kawai[38,39], J. Knödlseder[40], M. Kuss[6], J. Lande[3], S. Larsson[26,27,41*], L. Latronico[6], M. Lemoine-Goumard[30,31], F. Longo[8,9], F. Loparco[12,13], B. Lott[30,31], M. N. Lovellette[1], P. Lubrano[4,5], G. M. Madejski[3], A. Makeev[1,22], E. Massaro[42], M. N. Mazziotta[13], J. E. McEnery[19,43], P. F. Michelson[3], W. Mitthumsiri[3], T. Mizuno[33], A. A. Moiseev[20,43], C. Monte[12,13], M. E. Monzani[3], A. Morselli[44], I. V. Moskalenko[3], M. Mueller[3], S. Murgia[3], P. L. Nolan[3], J. P. Norris[45], E. Nuss[24], M. Ohno[46], T. Ohsugi[47], N. Omodei[3], E. Orlando[48], J. F. Ormes[45], M. Ozaki[46], J. H. Panetta[3], D. Parent[1,22,30,31], V. Pelassa[24], M. Pepe[4,5], M. Pesce-Rollins[6], F. Piron[24], T. A. Porter[3], S. Rainò[12,13], R. Rando[10,11], M. Razzano[6], A. Reimer[49,3], O. Reimer[49,3], S. Ritz[50], A. Y. Rodriguez[16], R. W. Romani[3], M. Roth[15], F. Ryde[51,27], H. F.-W. Sadrozinski[50], A. Sander[36], J. D. Scargle[52], C. Sgrò[6], M. S. Shaw[3], P. D. Smith[36], G. Spandre[6], P. Spinelli[12,13], J.-L. Starck[7], M. S. Strickman[1], D. J. Suson[53], H. Takahashi[47], T. Takahashi[46], T. Tanaka[3], J. B. Thayer[3], J. G. Thayer[3], D. J. Thompson[19], L. Tibaldo[10,11,7,54], D. F. Torres[35,16], G. Tosti[4,5*], A. Tramacere[3,55,56], Y. Uchiyama[3], T. L. Usher[3], V. Vasileiou[20,21], N. Vilchez[40], V. Vitale[44,57], A. P. Waite[3], E. Wallace[15], P. Wang[3], B. L. Winer[36], K. S. Wood[1], Z. Yang[26,27], T. Ylinen[51,58,27], M. Ziegler[50]

## ABSTRACT

This paper presents light curves as well as the first systematic characterization of variability of the 106 objects in the high-confidence Fermi Large Area Telescope (LAT) Bright AGN Sample (LBAS). Weekly light curves of this sample, obtained during the first 11 months of the Fermi survey (August 04, 2008 - July 04, 2009), are tested for variability, and their properties are quantified through autocorrelation function and structure function analysis. For the brightest sources, 3 or 4-day binned light curves are extracted in order to determine power density spectra (PDS) and to fit the temporal structure of major flares. More than 50% of the sources are found to be variable with high significance, where high states do not exceed 1/4 of the total observation range. Variation amplitudes are larger for flat spectrum radio quasars (FSRQs) and low/intermediate synchrotron frequency peaked (LSP/ISP) BL Lac objects. Autocorrelation time scales derived from weekly light curves vary from 4 to a dozen of weeks. Variable sources of the sample have weekly and 3 - 4 day bin light curves that can be described by $1/f^\alpha$ PDS, and show two kinds of gamma-ray variability: (1) rather constant baseline with sporadic flaring activity characterized by flatter PDS slopes resembling flickering and red-noise with occasional intermittence, and (2) - measured for a few blazars showing strong activity - complex and structured temporal profiles characterized by longer-term memory and steeper PDS slopes, reflecting a random-walk underlying mechanism. The average slope of the PDS of the brightest 22 FSRQs and of the 6 brightest BL Lacs is 1.5 and 1.7 respectively. The study of temporal profiles of well resolved flares observed in the 10 brightest LBAS sources shows that they generally have symmetric profiles and that their total duration vary between 10 and 100 days. Results presented here can assist in source class recognition for unidentified sources and can serve as reference for more detailed analysis of the brightest gamma-ray blazars.

*Subject headings:* gamma rays: observations — quasars: general — BL Lacertae objects: general — methods: statistical — galaxies: active

Corresponding authors: S. Ciprini, stefano.ciprini@pg.infn.it; S. Cutini, sarac@slac.stanford.edu; S. Larsson, stefan@astro.su.se; G. Tosti, Gino.Tosti@pg.infn.it.

[1] Space Science Division, Naval Research Laboratory, Washington, DC 20375, USA

[2] National Research Council Research Associate, National Academy of Sciences, Washington, DC 20001, USA

[3] W. W. Hansen Experimental Physics Laboratory, Kavli Institute for Particle Astrophysics and Cosmology, Department of Physics and SLAC National Accelerator Laboratory, Stanford University, Stanford, CA 94305, USA

[4] Istituto Nazionale di Fisica Nucleare, Sezione di Perugia, I-06123 Perugia, Italy

[5] Dipartimento di Fisica, Università degli Studi di Perugia, I-06123 Perugia, Italy

[6] Istituto Nazionale di Fisica Nucleare, Sezione di Pisa, I-56127 Pisa, Italy

[7] Laboratoire AIM, CEA-IRFU/CNRS/Université Paris Diderot, Service d'Astrophysique, CEA Saclay, 91191 Gif sur Yvette, France

[8] Istituto Nazionale di Fisica Nucleare, Sezione di Trieste, I-34127 Trieste, Italy

[9] Dipartimento di Fisica, Università di Trieste, I-34127 Trieste, Italy

[10] Istituto Nazionale di Fisica Nucleare, Sezione di Padova, I-35131 Padova, Italy

[11] Dipartimento di Fisica "G. Galilei", Università di Padova, I-35131 Padova, Italy

[12] Dipartimento di Fisica "M. Merlin" dell'Università e del Politecnico di Bari, I-70126 Bari, Italy

[13] Istituto Nazionale di Fisica Nucleare, Sezione di Bari, 70126 Bari, Italy

[14] Laboratoire Leprince-Ringuet, École polytechnique, CNRS/IN2P3, Palaiseau, France

[15] Department of Physics, University of Washington, Seattle, WA 98195-1560, USA

[16] Institut de Ciencies de l'Espai (IEEC-CSIC), Campus UAB, 08193 Barcelona, Spain

[17] INAF-Istituto di Astrofisica Spaziale e Fisica Cosmica, I-20133 Milano, Italy

[18] Agenzia Spaziale Italiana (ASI) Science Data Center, I-00044 Frascati (Roma), Italy

[19] NASA Goddard Space Flight Center, Greenbelt, MD 20771, USA

[20] Center for Research and Exploration in Space Science and Technology (CRESST) and NASA Goddard Space Flight Center, Greenbelt, MD 20771, USA

[21] Department of Physics and Center for Space Sciences and Technology, University of Maryland Baltimore County, Baltimore, MD 21250, USA

[22] George Mason University, Fairfax, VA 22030, USA

[24] Laboratoire de Physique Théorique et Astroparticules, Université Montpellier 2, CNRS/IN2P3, Montpellier, France

[25] Department of Physics and Astronomy, Sonoma State University, Rohnert Park, CA 94928-3609, USA

[26] Department of Physics, Stockholm University, AlbaNova, SE-106 91 Stockholm, Sweden

[27] The Oskar Klein Centre for Cosmoparticle Physics, AlbaNova, SE-106 91 Stockholm, Sweden

[28] Royal Swedish Academy of Sciences Research Fellow, funded by a grant from the K. A. Wallenberg Foundation

[29] Dipartimento di Fisica, Università di Udine and Istituto Nazionale di Fisica Nucleare, Sezione di Trieste, Gruppo Collegato di Udine, I-33100 Udine, Italy

[30] CNRS/IN2P3, Centre d'Études Nucléaires Bordeaux Gradignan, UMR 5797, Gradignan, 33175, France

[31] Université de Bordeaux, Centre d'Études Nucléaires Bordeaux Gradignan, UMR 5797, Gradignan, 33175, France

[32] Osservatorio Astronomico di Trieste, Istituto Nazionale di Astrofisica, I-34143 Trieste, Italy

[33] Department of Physical Sciences, Hiroshima University, Higashi-Hiroshima, Hiroshima 739-8526, Japan

[34] Center for Space Plasma and Aeronomic Research (CSPAR), University of Alabama in Huntsville, Huntsville, AL 35899, USA

[35] Institució Catalana de Recerca i Estudis Avançats (ICREA), Barcelona, Spain

[36] Department of Physics, Center for Cosmology and Astro-Particle Physics, The Ohio State University, Columbus, OH 43210, USA

[37] Research Institute for Science and Engineering, Waseda University, 3-4-1, Okubo, Shinjuku, Tokyo, 169-8555 Japan

[38] Department of Physics, Tokyo Institute of Technology, Meguro City, Tokyo 152-8551, Japan

[39] Cosmic Radiation Laboratory, Institute of Physical and Chemical Research (RIKEN), Wako, Saitama 351-0198, Japan

[40] Centre d'Étude Spatiale des Rayonnements, CNRS/UPS, BP 44346, F-30128 Toulouse Cedex 4, France

[41] Department of Astronomy, Stockholm University, SE-106 91 Stockholm, Sweden

[42] Physics Department, Università di Roma "La Sapienza", I-00185 Roma, Italy

[43] Department of Physics and Department of Astronomy, University of Maryland, College Park, MD 20742, USA

[44] Istituto Nazionale di Fisica Nucleare, Sezione di Roma

## 1. Introduction

The high energy emission and the erratic, rapid and large-amplitude variability observed in all accessible spectral regimes (radio-to-gamma-ray) are two of the main defining properties of blazars (e.g. Ulrich et al. 1997; Webb 2006). The entire non-thermal continuum is believed to originate mainly in a relativistic jet, pointing close to our line of sight. Studies of variability in different spectral bands and correlations of multi-waveband variability patterns allow us to shed light on the physical processes in action in blazars, such as particle acceleration and emission mechanisms, relativistic beaming, origin of flares and size, structure and location of the emission regions.

A proper understanding of the physical mechanisms responsible for variability is contingent upon a mathematical and statistical description of the phenomena. The study of variability is particularly important in gamma ray astronomy. First, it assists in detecting faint sources, discriminating between real point sources and background fluctuations. Second, correlated multi-wavelength variability helps to recognize and identify the correct radio/optical/X-ray source counterparts within the gamma-ray position error box. A characterization of stand-alone gamma-ray variability for unidentified sources can also support the recognition of the correct source class (Nolan et al. 2003).

Even though studied for many years, the details of blazar variability in various bands have not been consistently compared against each other. A major contribution to our current understanding of the blazar phenomenon has been provided by EGRET, which discovered blazars as the largest class of identified and variable gamma-ray sources, in the band above 100 MeV. EGRET blazars showed variations on timescales from days to months (for the sources observed in several viewing periods) and gamma-ray flares on short timescales (1–3 days) have been detected in PKS 0528+134, 3C 279, PKS 1406−076, PKS 1633+382, PKS 1622-297, 3C 454.3 (see, e.g. von Montigny et al. 1995; Wallace et al. 2000; Nolan et al. 2003; Thompson 2006). In some cases giant $\gamma$-ray outbursts were also found by EGRET, (as for 3C 279, Hartman et al. 2001), and very rapid variability at very high energy was resolved (for example by HESS in PKS 2155−304, Aharonian et al. 2007). However a complete characterization of the blazar gamma-ray variability was limited by statistics, number of the observations and by the EGRET pointed operating mode.

A new view on the gamma-ray variable sky is coming from the *Fermi* Large Area Telescope (LAT). Thanks to its large field of view (cover-

"Tor Vergata", I-00133 Roma, Italy

[45]Department of Physics and Astronomy, University of Denver, Denver, CO 80208, USA

[46]Institute of Space and Astronautical Science, JAXA, 3-1-1 Yoshinodai, Sagamihara, Kanagawa 229-8510, Japan

[47]Hiroshima Astrophysical Science Center, Hiroshima University, Higashi-Hiroshima, Hiroshima 739-8526, Japan

[48]Max-Planck Institut für extraterrestrische Physik, 85748 Garching, Germany

[49]Institut für Astro- und Teilchenphysik and Institut für Theoretische Physik, Leopold-Franzens-Universität Innsbruck, A-6020 Innsbruck, Austria

[50]Santa Cruz Institute for Particle Physics, Department of Physics and Department of Astronomy and Astrophysics, University of California at Santa Cruz, Santa Cruz, CA 95064, USA

[51]Department of Physics, Royal Institute of Technology (KTH), AlbaNova, SE-106 91 Stockholm, Sweden

[52]Space Sciences Division, NASA Ames Research Center, Moffett Field, CA 94035-1000, USA

[53]Department of Chemistry and Physics, Purdue University Calumet, Hammond, IN 46323-2094, USA

[54]Partially supported by the International Doctorate on Astroparticle Physics (IDAPP) program

[55]Consorzio Interuniversitario per la Fisica Spaziale (CIFS), I-10133 Torino, Italy

[56]INTEGRAL Science Data Centre, CH-1290 Versoix, Switzerland

[57]Dipartimento di Fisica, Università di Roma "Tor Vergata", I-00133 Roma, Italy

[58]School of Pure and Applied Natural Sciences, University of Kalmar, SE-391 82 Kalmar, Sweden

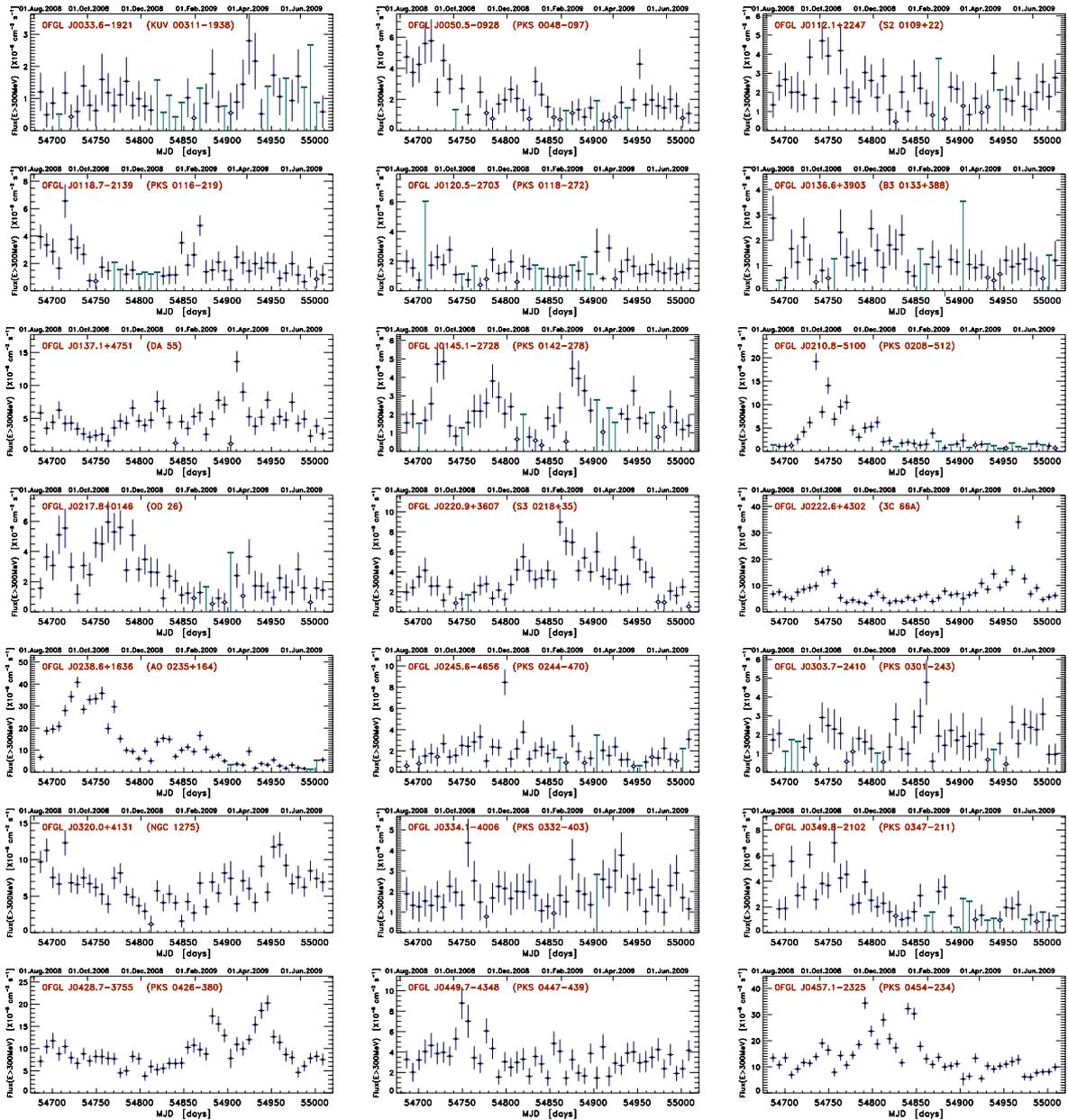

Fig. 1.— Light curves of the integrated flux ($E > 300$ MeV) measured in and averaged on weekly time bins obtained with the standard tool `gtlike`. In this picture (and continuations) all the 84 LBAS objects that are selected and used for a first temporal variability study are reported.

ing the 20% of the sky at any instant and the full sky in about 3 hours), improved effective area and sensitivity, and the all-sky operating mode, the LAT is, therefore, an unprecedented instrument to monitor the variability emission of blazars in the energy band 20 MeV to >300 GeV (see, e.g. Atwood et al. 2009; Abdo et al. 2009a, 2010a). A major result obtained by the *Fermi* LAT during the first three months of observations was the publication of the Bright Source List (0FGL, Abdo et al. 2009b), a list of 205 sources detected with a significance $> 10\sigma$. Of these, 106 sources located at $|b| > 10°$ have been associated with high confidence with known AGNs and constitute the LAT Bright AGN sample (LBAS). The LBAS sample include two radio galaxies (Cen A and NGC 1275) and 104 blazars, of which 58 are flat spectrum radio quasars (FSRQs), 42 are BL Lac objects, and 4 are blazars with uncertain classification (Abdo et al. 2009a).

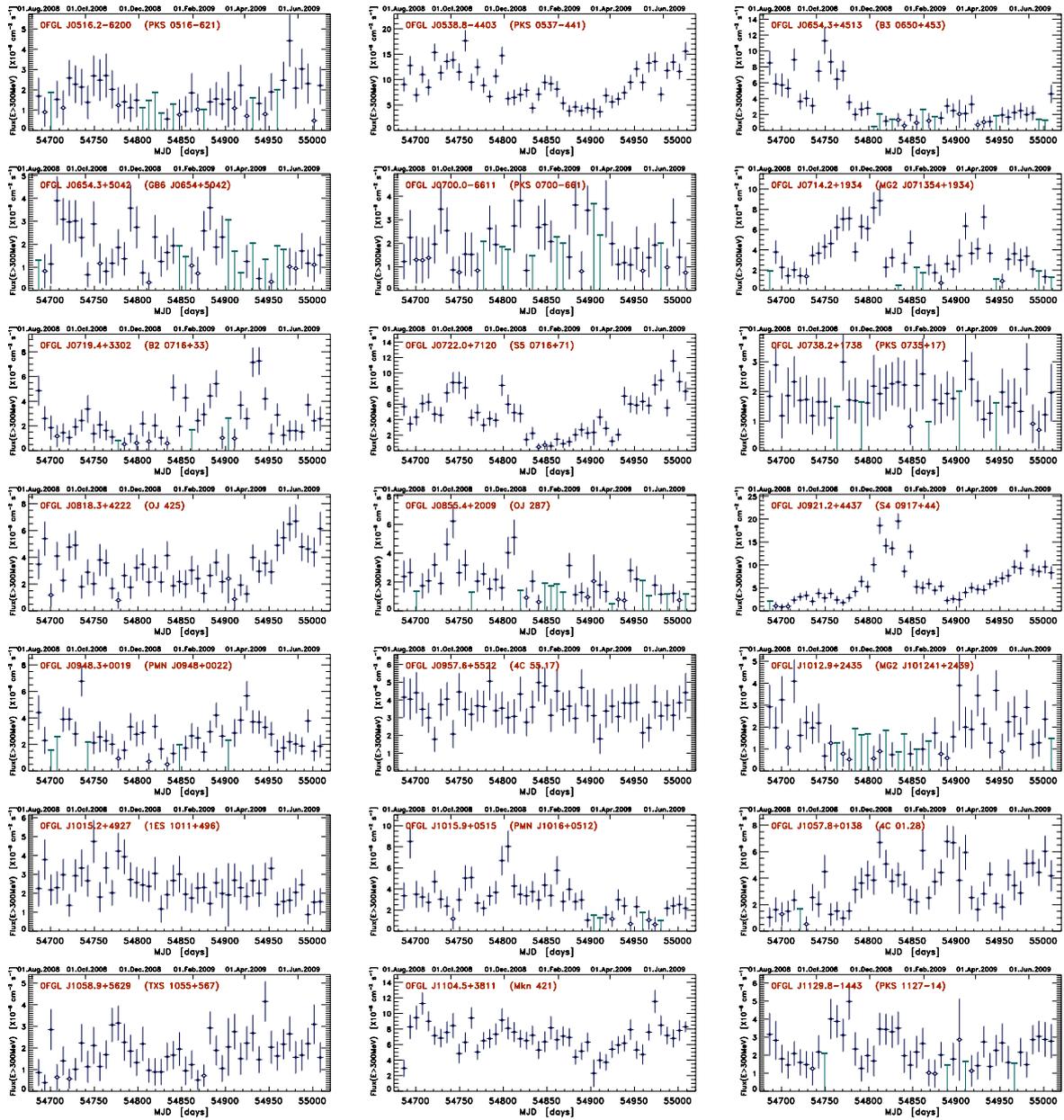

Fig. 2.— Continuation of Fig. 1.

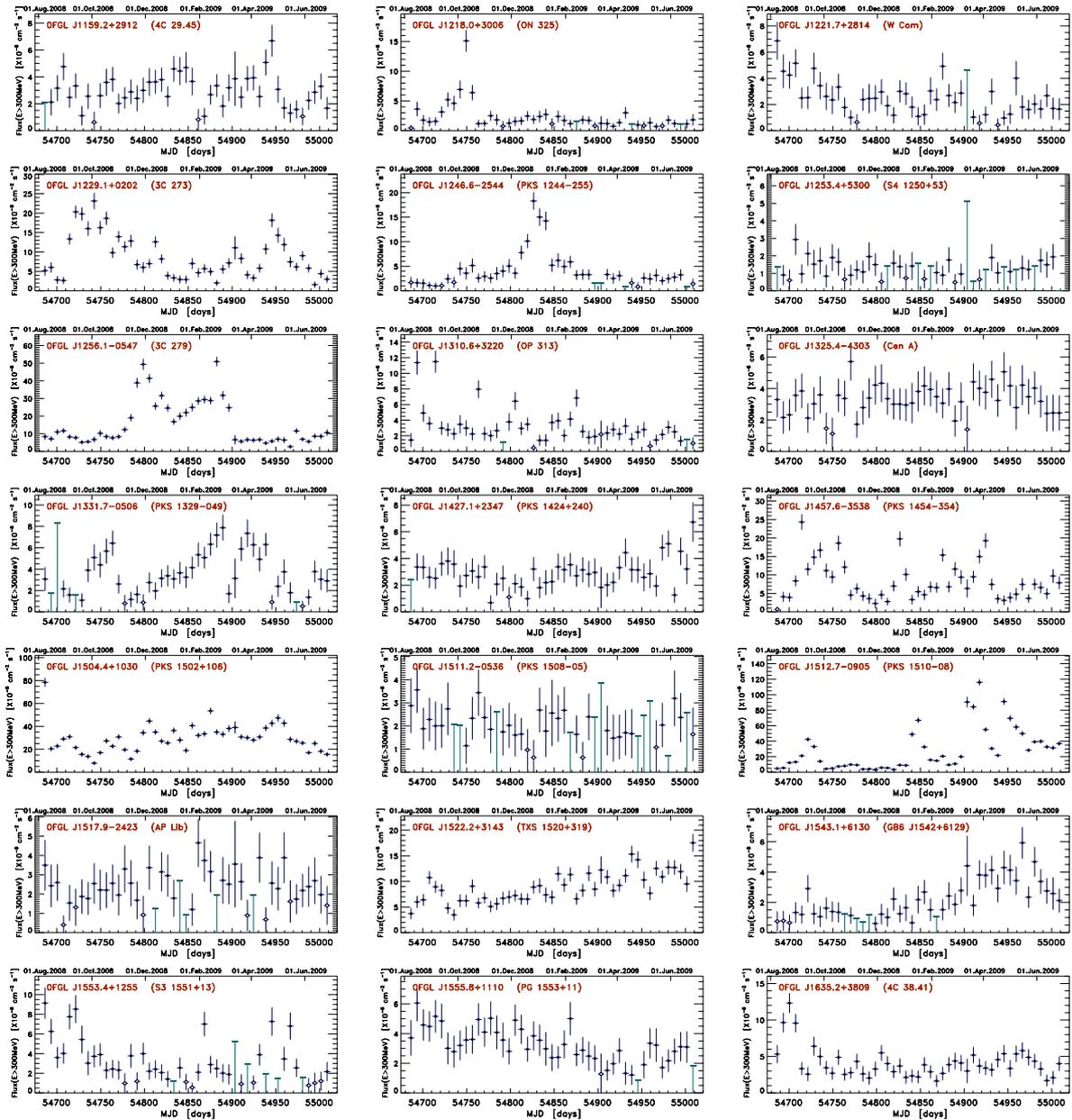

Fig. 3.— Continuation of Fig. 2.

This paper reports analysis results on the 11-month light curves of these 3-month selected bright AGNs, mostly blazars. Many of the light curves are, in fact, bright in the beginning of the considered 11-month period and then fade out. This does not represent a particular bias, as also in the 11-month detected source catalog (Abdo et al. 2010a, first year LAT catalog, 1FGL) on average these sources are among the brightest blazars. A parallel and detailed study of spectral properties on the same LBAS sample is presented in (Abdo et al. 2010b) while our analysis provides a temporal and flux variability analyses on the same sample. In Abdo et al. (2010b) the weekly gamma-ray spectral photon index is measured, in general, to vary in time only modestly (by <0.2–0.3) despite large variability of flux, and to vary only modestly within different blazar subclasses. In our paper significant gamma-ray flux variability for about half of the LBAS objects is reported and for 1/4 of the sources significant flares and outbursts are also found, evidencing a much stronger and violent variability of the flux than of the photon index.

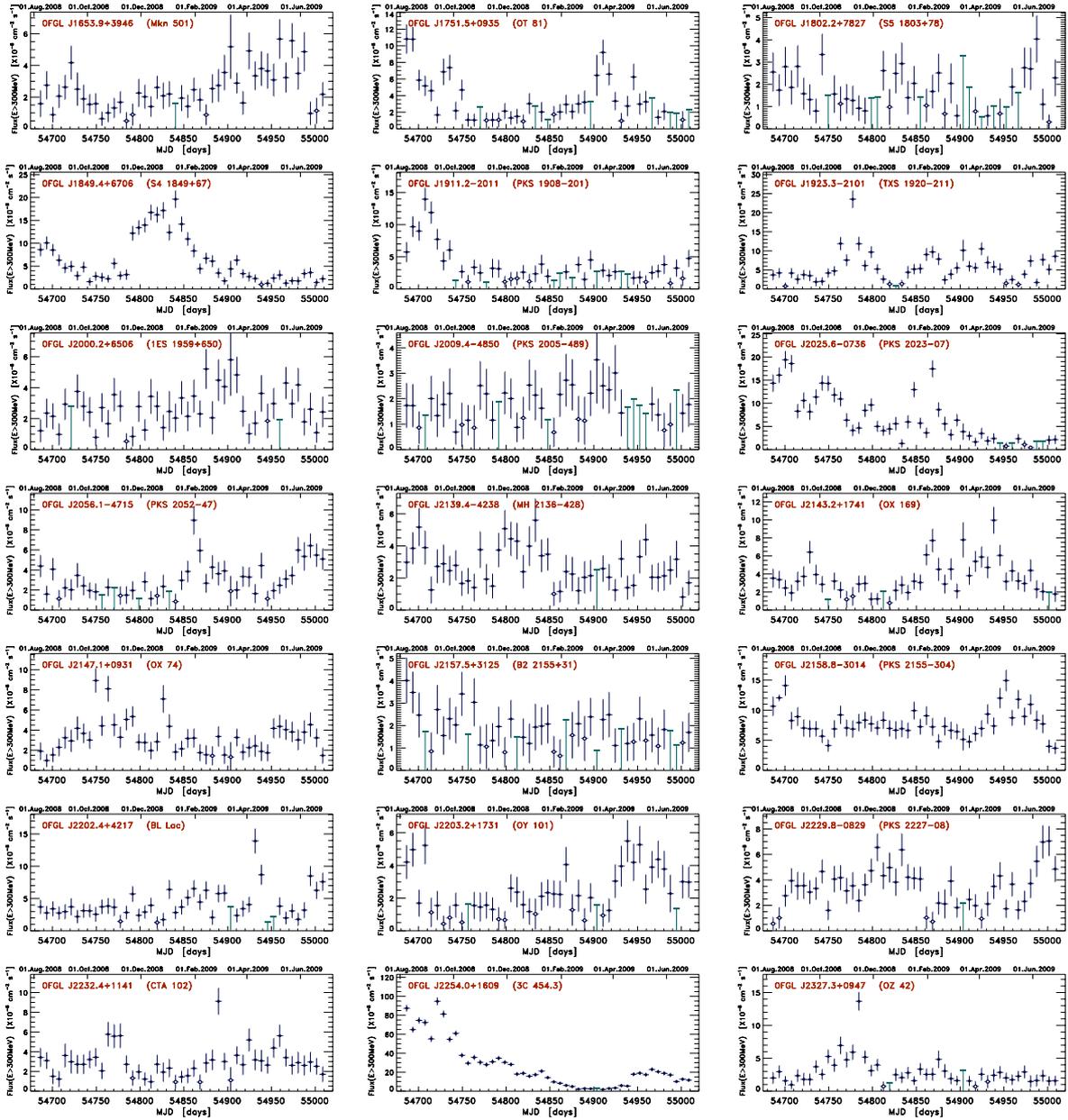

Fig. 4.— Continuation of Fig. 3.

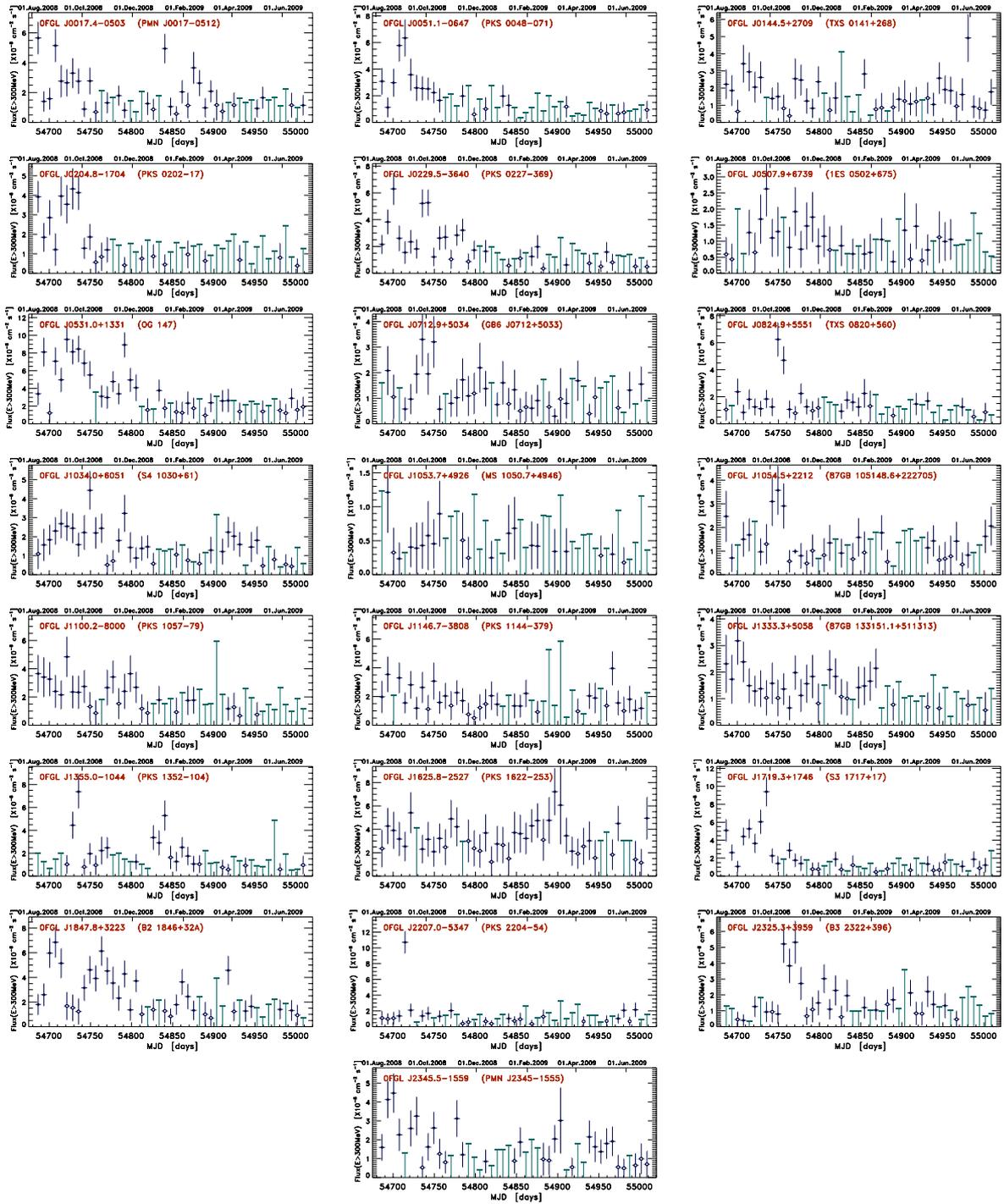

Fig. 5.— The 22 light curves that are excluded by the temporal analysis with the basis of the 60% - $TS \geq 4$ basis. Among these we note three peculiar light curves for the LSP blazars 0FGL J0531.0+1331, 0FGL J1719.3+1746, and 0FGL J2207.0−5347, showing strong flares and flux activity but during only a limited portion the 11-months observed.

In the following we use a $\Lambda$CDM (concordance) cosmology with values given within $1\sigma$ of the WMAP results (Komatsu et al. 2009), namely Hubble constant value $H_0 = 71$ km s$^{-1}$ Mpc$^{-1}$,

$\Omega_m = 0.27$, $\Omega_\Lambda = 0.73$.

In Section 2 a description of LAT observation and light curve extraction is reported, while in Section 3 results of variability search and amplitude quantification are presented. In Section 4 global variability properties of weekly-bin light curves are presented through autocorrelation and structure functions analysis. The analysis of the power spectral density and flare temporal profiles is presented in Section 5 and 6 respectively, using finer sampling (3 and 4 day bins) light curves for the brightest sources. Summary and conclusions are given in Section 7. Cross-correlation studies between the $\gamma$-ray and other bands (such as radio-mm and optical bands), as well as more detailed studies on periodicity search will be covered by other works based on the brightest sources where it is possible to obtain a finer-sampled flux light curves.

## 2. Observations with the LAT and LBAS light curves

The *Fermi*−LAT (Atwood et al. 2009; Abdo et al. 2009e) is a pair−conversion gamma−ray telescope sensitive to photon energies greater than 20 MeV. It consists of a tracker (composed of two sections, front and back, with different capabilities), a calorimeter and an anticoincidence system to reject the charged−particle background. The LAT has a large peak effective area ($\sim$ 8000 cm$^2$ for 1 GeV photons in the event class considered here), viewing $\approx$ 2.4 sr of the full sky with angular resolution (68% containment radius) better than $\approx 1°$ at $E = 1$ GeV.

Data used in this paper were collected during the first 11-month of nominal all-sky survey, from August 04, 2008 to July 04, 2009, (Modified Julian Day, MJD from 54682.655 to 55016.620).

In order to avoid background contamination from the bright Earth limb, time intervals where the Earth entered the LAT field of view (FoV) were excluded from this study (corresponding to a rocking angle larger than 47°). In addition, events that were reconstructed within 8° of the Earth limb were excluded from the analysis (corresponding to a zenith angle cut of 105°). Due to uncertainties in the current calibration, and the necessity of a trade-off between error accuracy and event statistics only photons belonging to the "Diffuse" class and with energies above 100 MeV were retained. This events analysis was performed with the standard *Fermi* LAT `ScienceTools` software package[1] (version v9r12) using in particular the tool `gtlike`, and using the first set of instrument response functions (IRFs) tuned with the flight data (P6_V3_DIFFUSE). In contrast to the preflight version, these IRFs take into account corrections for pile-up events. Since this is higher for lower energy photons, the measured photon index of a given source is about 0.1 higher (*i.e.* the spectrum is softer) with this IRF set as compared to the P6_V1_DIFFUSE one used previously in (Abdo et al. 2009b,a).

The light curves of all the LBAS sources were built using 7-day time intervals, for a total of 47 bins. For the brightest sources light curves were built using also time bins of 3 and 4 days (see Section 6.). For each time bin, the flux, photon index and test statistic (TS) of each source were determined, using the maximum-likelihood algorithm implemented in `gtlike`. The test statistic is defined as $TS = 2\Delta\log(\text{likelihood})$ between models with and without the source and it is a measure of the source significance (Mattox et al. 1996).

Photons were selected in a region of interest (RoI) of 7° in radius centered on the position of the source of interest. In the RoI analysis the sources were modeled as simple power-law ($F = kE^{-\Gamma}$). The isotropic background (the sum of residual instrumental background and extragalactic diffuse gamma-ray background) was modeled with a simple power-law. The GALPROP model version `gll_iem_v01.fit` (Strong et al. 2004a,b) was used for the Galac-tic diffuse emission, with both flux and spectral photon index left free in the model fit. All errors reported in the figures or quoted in the text are 1-$\sigma$ statistical errors. The estimated systematic uncertainty on the flux is 10% at 100 MeV, 5% at 500 MeV and 20% at 10 GeV.

---
[1] http://fermi.gsfc.nasa.gov/ssc/data/analysis/documentation/Cicerone/

For the 106 weekly light curves of LBAS sources analyzed in Section 3 and 4 the flux in each time bin is reported for the energy band $E > 300$ MeV. This is the best band for reporting the flux because it is the band for which we have the highest signal to noise ratio for each source. The 3-day and 4-day bin light curves of the brightest sources analyzed in Sections 5 and 6 are extracted using the F($E > 100$ MeV) flux.

## 3. Variability search and amplitude in weekly light curves

The following variability analysis was performed using the weekly light curves reported in Figures 1 – 5. We used F($E > 300$ MeV) (hereafter $F_{300}$) in order to enable the comparison of the observed variability characteristics for the different sources.

Because of the intrinsically variable nature of blazars for several sources we were not able to obtain a highly significant ($TS > 25$) estimate of the flux for all the 47 weeks. Therefore in building the light curves we followed the same approach described in Abdo et al. (2009b). For each time bin we keep the best fit value of the flux and its estimated error and when the $TS < 1$ we computed the $1\sigma$ upper limit.

We investigated whether a source had significant variations using a simple $\chi^2$ test.

$$\chi^2 = \sum_{i=1}^{N_p} \frac{(F_i - \langle F_i \rangle^2)}{(\sigma_i^2 + \sigma_{syst}^2)} \quad (1)$$

where $F_i$ are the $F_{300}$ fluxes of each source on each bin and $\sigma_i$ is the statistical uncertainty to which we added in quadrature $\sigma_{syst} = 0.03\langle F_i \rangle$ as an estimate of the systematic error (Abdo et al. 2009e,a); $N_p$ is the number of points in each light curve having a $TS \geq 4 (\sim 2\sigma)$ and $\langle F_i \rangle$ is the unweighted mean of the flux. This test was applied to light curves containing only flux values with $TS \geq 4 (\sim 2\sigma)$ and excluding upper limits and fluxes with $\sigma_i/F_i > 0.5$ (see Figure 6). Figure 7 shows the distribution of the "coverage", that is the detection fraction of the total period of observation after this cut on the TS. The weekly light curves for the 84 LBAS objects for which this fraction is $> 60\%$, i.e. having at least 28 detections with $TS \geq 4$, are shown in Figures 1, 2, 3 and 4. The light curves of lower quality - corresponding to the remaining 22 are shown in Figure 5.

We also quantify the variability amplitude of all the LBAS sources, using the "normalized excess variance" (Nandra et al. 1997; Edelson et al. 2002). This estimator is defined by:

$$\sigma_{NXS}^2 = \frac{S^2 - \langle \sigma_{err}^2 \rangle}{\langle F_i \rangle^2} \quad (2)$$

where $S^2$ is the variance of the light curve and $\sigma_{err}^2 = \sigma_i^2 + \sigma_{sys}^2$. The error in $\sigma_{NXS}^2$ was evaluated according to the prescription of Vaughan et al. (2003).

The results of this analysis are reported in Table 1: the first column lists the bright source list (0FGL list) name, column 2 the other source name, column 3 the optical class. In the fourth column we report the spectral energy distribution (SED) class, based on the peak frequency of the synchrotron component ($\nu_p^S$) of the broadband SED following the scheme outlined by Abdo et al. (2010c) which is an extension of the classification system introduced by (Padovani & Giommi 1995) for BL Lacs. In this scheme we have: low synchrotron-peaked (LSP, for $\nu_{peak}^S < 10^{14}$ Hz), intermediate synchrotron-peaked (ISP, for $10^{14}$ Hz $< \nu_{peak}^S < 10^{15}$ Hz) and high synchrotron-peaked (HSP, for $\nu_{peak}^S > 10^{15}$ Hz) blazars. Data listed in columns 5 – 13 are the redshift, $N_p$, the mean flux the standard deviation of each light curve, the peak flux and error, the variability probability of $\chi^2$ (for $N_p - 1$ degrees of freedom), the normalized excess variance and error. Negative values of $\sigma^2_{NXS}$ indicate absence or very small variability and/or slightly overestimated errors.

The large majority of sources (74) belong to the LSP class, which includes all FSRQs (58) and several BL Lacs (16), while both ISP and HSP classes have each 13 BL Lacs sources. There are also 6 objects which cannot be well classified for paucity of data or because they are peculiar AGNs defined commonly as radio galaxies, such as NGC 1275 (Per A) and Cen A.

On the basis of the $\chi^2$ test, variability was detected in 68 out of 106 LBAS sources with a significance higher than 99% (column 11 in Table 1). Note, however, that as demonstrated by Figure 19 in Abdo et al. (2010e) the $\chi^2$ has a strong dependence on the statistical flux uncertainties. For the fainter sources this leads to a reduction of the $\chi^2$ for a given fractional flux variation and then a source can be considered significantly variable only if it is both intrinsically variable and sufficiently bright. Therefore, fainter sources can appear less variable than brighter sources simply because we cannot measure their variability.

In Abdo et al. (2009a) 56 sources were flagged as variable based on the results of a $\chi^2$ test applied to weekly light curves covering the first three months of operation. To compare our results with those reported in Abdo et al. (2009a), we divided the light curves in four consecutive segments having a duration of about twelve weeks, and the $\chi^2$ test was applied to each of them. 42 sources were found variable with a significance higher than 99% during the first light curve segment (corresponding to about the same time interval analyzed in Abdo et al. (2009b,a), 28 in the second, 23 in the the third and 19 in the last. The difference in the number of variable sources in the first segment with respect to Abdo et al. (2009a) results, can be explained taking into account that in the Abdo et al. (2009a) all light curve data points, including those with TS <4, were considered in the calculation of the $\chi^2$ and the likelihood analysis was performed with a different combination of IRFs and diffuse models. The decreasing number of variable sources revealed in the four time intervals is a selection effect. We are using the BSL sample, so there are disproportionally more objects which happened to flare up at the beginning of the interval and then faded. However this is illustrative of one of the distinctive aspect of the intrinsic characteristics of the blazars' variability; alternate periods of flaring and low activity states. However the total period of our observations is still too short to allow an estimation of the duty cycle of the blazar variability in the gamma-ray energy range.

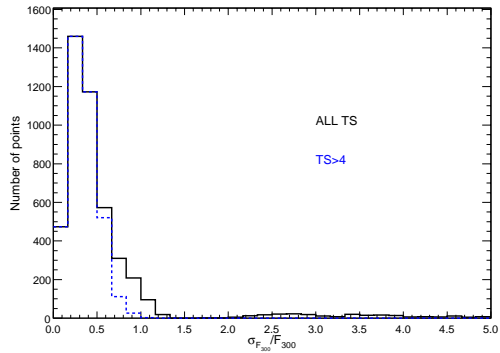

Fig. 6.— Distribution of the relative flux errors $\sigma_{F_{300}}/F_{300}$ for all the 106 LBAS light curves and all the data points. The larger values of the relative error in the distribution labeled "All TS" are due to the counting of upper limits.

Figure 8 shows the distribution of the peak $F_{300}$ ($F_M$) values for LSP, ISP and HSP. It can be noted that only a few LSP were detected in exceptionally bright states with a flux $F_M > 2 \times 10^{-7}$ ph cm$^{-2}$ s$^{-1}$.

Figure 9 and Figure 10 show the distributions of $\sigma^2_{NXS}$ and of the ratio between the highest measured flux to the mean $F_M/\langle F \rangle$ for the above three SED classes. These figures were obtained using only the 84 light curves with a coverage $\geq 60\%$. Variability amplitude of LSPs is generally larger than for ISPs and HSPs with the remarkable exception of the HSP source 0FGL J1218.0+3006 (ON 325 also known as Ton 605) which has the higher values of $F_M/\langle F \rangle$ among the LBAS sources. This source was always close to the detection limit on a week time scale, but a strong flare was observed during October 10 – 15, 2009. This shows that although HSP seems to have, on average, a variability amplitude smaller than those observed in LSP, episodic large flaring activity can be observed also for this subclass of blazars.

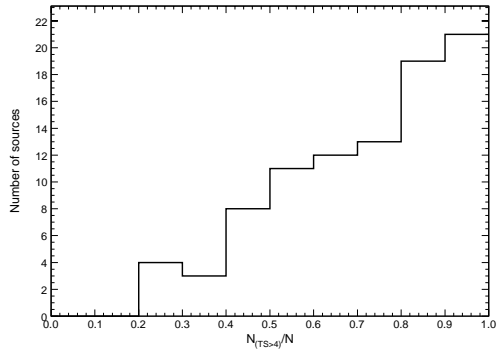

Fig. 7.— Distribution of the coverage fraction of the observation period of each light curve.

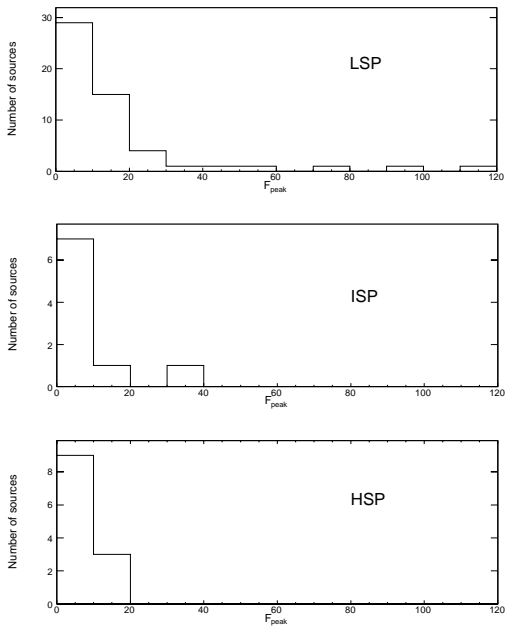

Fig. 8.— Distribution of $F_M$ for the LSP, ISP and HSP light curves with a coverage factor $\geq 60\%$.

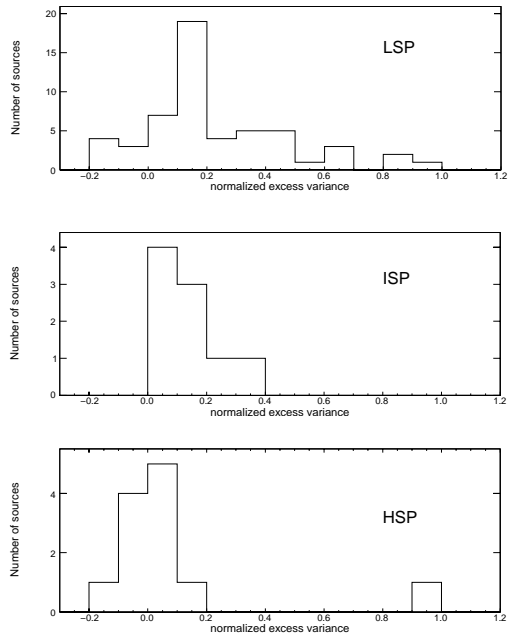

Fig. 9.— Distribution of the excess variance for the LSP, ISP and HSP light curves with a coverage factor $\geq 60\%$.

Among the sources with a coverage $< 60\%$ (22 sources), three sources have $F_M \sim 10^{-7}$ ph cm$^{-2}$ s$^{-1}$. 0FGL J2207.0−5347 (PKS 2204−54) has a light curve dominated by a short and intense flare detected during September 3 – 8, 2008; 0FGL J1719.3+1746 (PKS 1717+177) was mainly active during September 2008; 0FGL J0531.0+1331 (PKS 0528+134), one of the most active source during the EGRET era, was in a relative bright state during September-November 2008, with two flaring episodes, then it decreased to a flux close to the Fermi-LAT detection threshold on a week time scale.

To obtain an estimate of the time spent by each source in a bright state we evaluate the num-ber of time bins ($N_b$) for which $(F_i - \sigma_i) > (\langle F \rangle + 1.5 \times S)$. The distribution of the ratio $N_b/N$ in percent is reported in Figure 11. We see that high states exceeding one fourth of the duration of entire observation window are absent and that a very high number of sources were bright over a time interval shorter than the 5% of the total observation time. This distribution can be approximately described by a power law ($N_S = (332 \pm 72) \times (100 * N_b/N)^{-1.54 \pm 0.15}$).

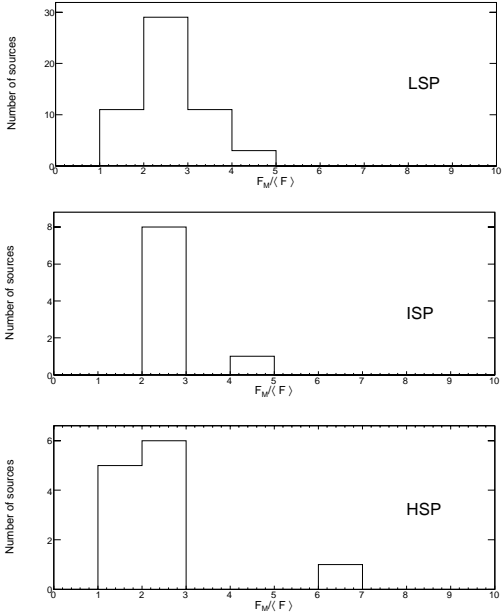

Fig. 10.— Distribution of $F_M/\langle F \rangle$ for the LSP, ISP and HSP light curves with a coverage factor $\geq 60\%$.

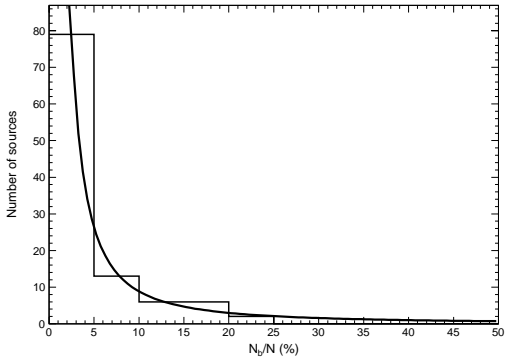

Fig. 11.— Distribution of $N_b/N$. This distribution could be described by a power law: $N_S = (332 \pm 72) \times (100 * N_b/N)^{-1.54 \pm 0.15}$.

TABLE 1
VARIABILITY INDICES AND AMPLITUDES.

| 0FGL | Other Name | Optical Class | SED Class | $z$ | $N_p$ | $\langle F_{300}\rangle$[a] | $S$[a] | $F_M$[a] | $\sigma_M$[a] | Prob | $\sigma_{NXS}^2$ | err($\sigma_{NXS}^2$) |
|---|---|---|---|---|---|---|---|---|---|---|---|---|
| J0017.4−0503 | PMN J0017−0512 | FSRQ | LSP | 0.227 | 24 | 2.30 | 1.37 | 5.65 | 1.07 | >99.0 | 0.24 | 0.08 |
| J0033.6−1921 | KUV 00311−1938 | BLLac | HSP | 0.610 | 31 | 1.12 | 0.52 | 2.80 | 1.04 | 12.0 | −0.09 | 0.08 |
| J0050.5−0928 | PKS 0048−097 | BLLac | ISP | >0.30 | 32 | 2.51 | 1.32 | 5.77 | 1.37 | >99.0 | 0.16 | 0.06 |
| J0051.1−0647 | PKS 0048−071 | FSRQ | LSP | 1.975 | 14 | 2.84 | 1.47 | 6.34 | 1.38 | >99.0 | 0.17 | 0.09 |
| J0112.1+2247 | S2 0109+22 | BLLac | ISP | >0.23 | 39 | 2.20 | 0.88 | 4.70 | 0.97 | 69.8 | 0.02 | 0.04 |
| J0118.7−2139 | PKS 0116−219 | FSRQ | LSP | 1.165 | 39 | 2.07 | 1.19 | 6.56 | 1.21 | >99.0 | 0.19 | 0.06 |
| J0120.5−2703 | PKS 0118−272 | BLLac | LSP | 0.557 | 34 | 1.52 | 0.55 | 2.89 | 0.91 | 1.8 | −0.11 | 0.06 |
| J0136.6+3903 | B3 0133+388 | BLLac | HSP | ⋯ | 34 | 1.29 | 0.55 | 2.87 | 0.89 | 11.8 | −0.07 | 0.06 |
| J0137.1+4751 | DA 55 | FSRQ | LSP | 0.859 | 45 | 4.89 | 2.13 | 13.62 | 1.56 | >99.0 | 0.13 | 0.03 |
| J0144.5+2709 | TXS 0141+268 | BLLac | ISP | ⋯ | 26 | 1.98 | 0.91 | 4.92 | 1.38 | 72.3 | 0.03 | 0.06 |
| J0145.1−2728 | PKS 0142−278 | FSRQ | LSP | 1.148 | 33 | 2.32 | 1.05 | 4.85 | 0.96 | >99.0 | 0.10 | 0.05 |
| J0204.8−1704 | PKS 0202−17 | FSRQ | LSP | 1.740 | 11 | 2.74 | 1.22 | 4.33 | 1.05 | >99.0 | 0.12 | 0.08 |
| J0210.8−5100 | PKS 0208−512 | FSRQ | LSP | 1.003 | 30 | 4.43 | 4.22 | 19.19 | 1.79 | >99.0 | 0.87 | 0.09 |
| J0217.8+0146 | OD 26 | FSRQ | LSP | 1.715 | 40 | 2.70 | 1.47 | 5.95 | 1.36 | >99.0 | 0.19 | 0.05 |
| J0220.9+3607 | S3 0218+35 | FSRQ | LSP | 0.944 | 42 | 3.55 | 1.71 | 8.98 | 1.37 | >99.0 | 0.15 | 0.04 |
| J0222.6+4302 | 3C 66A | BLLac | ISP | 0.444 | 47 | 7.92 | 5.05 | 34.06 | 2.52 | >99.0 | 0.38 | 0.03 |
| J0229.5−3640 | PKS 0227−369 | FSRQ | LSP | 2.115 | 20 | 2.62 | 1.45 | 6.30 | 1.18 | >99.0 | 0.22 | 0.07 |
| J0238.6+1636 | AO 0235+164 | BLLac | LSP | 0.940 | 44 | 13.19 | 10.73 | 40.78 | 2.42 | >99.0 | 0.66 | 0.03 |
| J0245.6−4656 | PKS 0244−470 | Un | ⋯ | ⋯ | 35 | 2.20 | 1.28 | 8.46 | 1.18 | >99.0 | 0.20 | 0.07 |
| J0303.7−2410 | PKS 0301−243 | BLLac | HSP | 0.260 | 36 | 1.95 | 0.78 | 4.78 | 1.19 | 43.3 | −0.01 | 0.04 |
| J0320.0+4131 | NGC 1275 | RG | ⋯ | 0.018 | 46 | 6.54 | 2.42 | 12.30 | 1.71 | >99.0 | 0.09 | 0.02 |
| J0334.1−4006 | PKS 0332−403 | BLLac | LSP | ⋯ | 44 | 1.98 | 0.72 | 4.36 | 1.16 | 15.1 | −0.04 | 0.04 |
| J0349.8−2102 | PKS 0347−211 | FSRQ | LSP | 2.944 | 33 | 2.86 | 1.51 | 7.00 | 1.55 | >99.0 | 0.18 | 0.05 |
| J0428.7−3755 | PKS 0426−380 | BLLac | LSP | 1.112 | 47 | 9.19 | 3.58 | 20.24 | 1.74 | >99.0 | 0.13 | 0.02 |
| J0449.7−4348 | PKS 0447−439 | BLLac | HSP | 0.205 | 47 | 3.40 | 1.44 | 8.79 | 1.55 | >99.0 | 0.09 | 0.03 |
| J0457.1−2325 | PKS 0454−234 | FSRQ | LSP | 1.003 | 47 | 13.56 | 6.78 | 34.39 | 2.23 | >99.0 | 0.24 | 0.02 |
| J0507.9+6739 | 1ES 0502+675 | BLLac | HSP | 0.416 | 23 | 1.14 | 0.50 | 2.63 | 0.85 | 6.3 | −0.14 | 0.10 |
| J0516.2−6200 | PKS 0516−621 | Un | ⋯ | ⋯ | 29 | 1.95 | 0.74 | 4.42 | 1.28 | 9.1 | −0.08 | 0.06 |
| J0531.0+1331 | PKS 0528+134 | FSRQ | LSP | 2.070 | 22 | 5.04 | 2.34 | 9.54 | 1.46 | >99.0 | 0.15 | 0.05 |
| J0538.8−4403 | PKS 0537−441 | BLLac | LSP | 0.892 | 47 | 9.23 | 3.58 | 17.65 | 2.06 | >99.0 | 0.13 | 0.02 |
| J0654.3+4513 | B3 0650+453 | FSRQ | LSP | 0.933 | 32 | 4.10 | 2.63 | 11.29 | 1.66 | >99.0 | 0.35 | 0.06 |
| J0654.3+5042 | GB6 J0654+5042 | Un | ⋯ | ⋯ | 28 | 2.00 | 0.94 | 3.89 | 1.06 | 91.4 | 0.06 | 0.06 |
| J0700.0−6611 | PKS 0700−661 | Un | ⋯ | ⋯ | 29 | 2.12 | 0.87 | 3.81 | 1.15 | 39.4 | −0.02 | 0.05 |
| J0712.9+5034 | GB6 J0712+5033 | ⋯ | ⋯ | ⋯ | 24 | 1.40 | 0.76 | 3.30 | 1.04 | 69.3 | 0.03 | 0.09 |
| J0714.2+1934 | MG2 J071354+1934 | FSRQ | LSP | 0.534 | 37 | 3.98 | 1.97 | 8.85 | 1.48 | >99.0 | 0.17 | 0.04 |
| J0719.4+3302 | B2 0716+33 | FSRQ | LSP | 0.779 | 37 | 2.78 | 1.59 | 7.26 | 1.11 | >99.0 | 0.23 | 0.06 |
| J0722.0+7120 | S5 0716+71 | BLLac | ISP | 0.310 | 45 | 4.94 | 2.63 | 11.56 | 1.44 | >99.0 | 0.24 | 0.03 |
| J0738.2+1738 | PKS 0735+17 | BLLac | LSP | 0.424 | 39 | 1.89 | 0.51 | 3.03 | 0.98 | 0.0 | −0.11 | 0.04 |
| J0818.3+4222 | OJ 425 | BLLac | LSP | 0.530 | 43 | 3.36 | 1.40 | 6.69 | 1.23 | >99.0 | 0.09 | 0.03 |
| J0824.9+5551 | TXS 0820+560 | FSRQ | LSP | 1.417 | 20 | 1.91 | 1.28 | 6.23 | 1.25 | >99.0 | 0.32 | 0.11 |
| J0855.4+2009 | OJ 287 | BLLac | LSP | 0.306 | 28 | 2.48 | 1.22 | 6.22 | 1.17 | >99.0 | 0.12 | 0.06 |
| J0921.2+4437 | S4 0917+44 | FSRQ | LSP | 2.190 | 44 | 6.56 | 4.32 | 19.52 | 1.70 | >99.0 | 0.41 | 0.04 |
| J0948.3+0019 | PMN J0948+0022 | FSRQ | LSP | 0.585 | 39 | 2.83 | 1.16 | 6.77 | 1.13 | >99.0 | 0.07 | 0.04 |
| J0957.6+5522 | 4C 55.17 | FSRQ | LSP | 0.896 | 47 | 3.55 | 0.76 | 5.05 | 1.07 | 2.1 | −0.03 | 0.02 |
| J1012.9+2435 | MG2 J101241+2439 | FSRQ | LSP | 1.805 | 28 | 2.11 | 0.93 | 4.09 | 0.98 | 71.7 | 0.02 | 0.05 |
| J1015.2+4927 | 1ES 1011+496 | BLLac | HSP | 0.212 | 47 | 2.40 | 0.79 | 4.74 | 1.12 | 36.4 | −0.01 | 0.03 |
| J1015.9+0515 | PMN J1016+0512 | FSRQ | LSP | 1.713 | 38 | 3.54 | 1.57 | 8.50 | 1.58 | >99.0 | 0.12 | 0.04 |
| J1034.0+6051 | S4 1030+61 | FSRQ | LSP | 1.401 | 24 | 2.00 | 0.74 | 4.44 | 1.01 | 52.4 | 0.00 | 0.04 |

TABLE 1—*Continued*

| 0FGL | Other Name | Optical Class | SED Class | $z$ | $N_P$ | $\langle F_{300}\rangle$[a] | $S$[a] | $F_M$[a] | $\sigma_M$[a] | Prob | $\sigma^2_{NXS}$ | err($\sigma^2_{NXS}$) |
|---|---|---|---|---|---|---|---|---|---|---|---|---|
| J1053.7+4926 | MS 1050.7+4946 | BLLac | ISP | 0.140 | 16 | 0.50 | 0.25 | 1.21 | 0.59 | 3.3 | −0.33 | 0.21 |
| J1054.5+2212 | 87GB 105148.6+222705 | BLLac | ISP | ⋯ | 20 | 1.64 | 0.79 | 3.57 | 1.03 | 71.2 | 0.03 | 0.08 |
| J1057.8+0138 | 4C 01.28 | FSRQ | LSP | 0.888 | 44 | 3.50 | 1.62 | 6.77 | 1.21 | >99.0 | 0.12 | 0.04 |
| J1058.9+5629 | TXS 1055+567 | BLLac | ISP | 0.143 | 44 | 1.83 | 0.79 | 4.15 | 0.92 | 78.7 | 0.03 | 0.04 |
| J1100.2−8000 | PKS 1057−79 | BLLac | LSP | 0.569 | 18 | 2.69 | 0.86 | 4.84 | 1.42 | 9.2 | −0.08 | 0.06 |
| J1104.5+3811 | Mkn 421 | BLLac | HSP | 0.030 | 47 | 6.84 | 1.88 | 11.54 | 1.46 | >99.0 | 0.04 | 0.01 |
| J1129.8−1443 | PKS 1127−14 | FSRQ | LSP | 1.184 | 38 | 2.44 | 0.86 | 4.98 | 0.99 | 50.3 | 0.00 | 0.03 |
| J1146.7−3808 | PKS 1144−379 | FSRQ | LSP | 1.048 | 24 | 2.06 | 0.77 | 3.96 | 1.17 | 14.4 | −0.06 | 0.06 |
| J1159.2+2912 | 4C 29.45 | FSRQ | LSP | 0.729 | 43 | 3.06 | 1.14 | 6.68 | 1.17 | 98.1 | 0.05 | 0.03 |
| J1218.0+3006 | ON 325 | BLLac | HSP | 0.130 | 38 | 2.51 | 2.51 | 15.11 | 1.82 | >99.0 | 0.90 | 0.12 |
| J1221.7+2814 | W Com | BLLac | ISP | 0.102 | 43 | 2.58 | 1.27 | 6.86 | 1.44 | >99.0 | 0.12 | 0.05 |
| J1229.1+0202 | 3C 273 | FSRQ | LSP | 0.158 | 47 | 8.68 | 5.47 | 23.11 | 2.07 | >99.0 | 0.38 | 0.03 |
| J1246.6−2544 | PKS 1244−255 | FSRQ | LSP | 0.635 | 37 | 4.60 | 3.81 | 18.28 | 1.75 | >99.0 | 0.64 | 0.07 |
| J1253.4+5300 | S4 1250+53 | BLLac | LSP | ⋯ | 29 | 1.45 | 0.47 | 2.93 | 0.89 | 0.9 | −0.12 | 0.06 |
| J1256.1−0548 | 3C 279 | FSRQ | LSP | 0.536 | 47 | 15.69 | 12.31 | 50.91 | 2.61 | >99.0 | 0.62 | 0.03 |
| J1310.6+3220 | OP 313 | FSRQ | LSP | 0.997 | 41 | 3.38 | 2.31 | 11.50 | 1.39 | >99.0 | 0.40 | 0.06 |
| J1325.4−4303 | Cen A | RG | ⋯ | 0.002 | 44 | 3.41 | 0.82 | 5.71 | 1.20 | 0.8 | −0.05 | 0.02 |
| J1331.7−0506 | PKS 1329−049 | FSRQ | LSP | 2.150 | 39 | 3.86 | 1.86 | 7.88 | 1.20 | >99.0 | 0.16 | 0.04 |
| J1333.3+5058 | 87GB 133151.1+511313 | FSRQ | LSP | 1.362 | 20 | 1.72 | 0.53 | 3.18 | 0.89 | 4.8 | −0.09 | 0.06 |
| J1355.0−1044 | PKS 1352−104 | FSRQ | LSP | 0.330 | 13 | 2.92 | 1.76 | 7.38 | 1.45 | >99.0 | 0.27 | 0.11 |
| J1427.1+2347 | PKS 1424+240 | BLLac | ISP | ⋯ | 45 | 2.91 | 1.08 | 6.73 | 1.47 | 94.0 | 0.04 | 0.03 |
| J1457.6−3538 | PKS 1454−354 | FSRQ | LSP | 1.424 | 46 | 8.49 | 5.16 | 24.30 | 2.08 | >99.0 | 0.34 | 0.03 |
| J1504.4+1030 | PKS 1502+106 | FSRQ | LSP | 1.839 | 47 | 29.57 | 12.15 | 78.44 | 3.79 | >99.0 | 0.17 | 0.01 |
| J1511.2−0536 | PKS 1508−05 | FSRQ | LSP | 1.185 | 31 | 2.17 | 0.58 | 3.56 | 0.99 | 0.0 | −0.14 | 0.05 |
| J1512.7−0905 | PKS 1510−08 | FSRQ | LSP | 0.360 | 47 | 28.67 | 27.21 | 115.94 | 3.82 | >99.0 | 0.91 | 0.02 |
| J1517.9−2423 | AP Lib | BLLac | LSP | 0.048 | 35 | 2.62 | 0.76 | 4.65 | 1.24 | 0.5 | −0.09 | 0.04 |
| J1522.2+3143 | TXS 1520+319 | FSRQ | LSP | 1.487 | 47 | 8.88 | 3.01 | 17.53 | 1.69 | >99.0 | 0.09 | 0.01 |
| J1543.1+6130 | GB6 J1542+6129 | BLLac | ISP | ⋯ | 39 | 2.39 | 1.26 | 5.93 | 1.03 | >99.0 | 0.16 | 0.05 |
| J1553.4+1255 | S3 1551+13 | FSRQ | LSP | 1.308 | 32 | 3.92 | 2.13 | 9.14 | 1.57 | >99.0 | 0.22 | 0.05 |
| J1555.8+1110 | PG 1553+11 | BLLac | HSP | >0.09 | 44 | 3.31 | 1.12 | 6.05 | 1.21 | 93.2 | 0.03 | 0.02 |
| J1625.8−2527 | PKS 1622−253 | FSRQ | LSP | 0.786 | 26 | 3.90 | 1.20 | 7.21 | 2.00 | 3.3 | −0.08 | 0.05 |
| J1635.2+3809 | 4C 38.41 | FSRQ | LSP | 1.814 | 47 | 4.09 | 2.06 | 12.32 | 1.34 | >99.0 | 0.19 | 0.04 |
| J1653.9+3946 | Mkn 501 | BLLac | HSP | 0.033 | 42 | 2.61 | 1.27 | 5.67 | 1.23 | >99.0 | 0.12 | 0.05 |
| J1719.3+1746 | S3 1717+17 | BLLac | LSP | 0.137 | 18 | 3.04 | 2.18 | 9.40 | 1.61 | >99.0 | 0.44 | 0.11 |
| J1751.5+0935 | OT 81 | BLLac | LSP | 0.322 | 33 | 3.96 | 2.72 | 10.82 | 1.73 | >99.0 | 0.38 | 0.07 |
| J1802.2+7827 | S5 1803+78 | BLLac | LSP | 0.680 | 30 | 1.94 | 0.85 | 4.04 | 1.05 | 77.3 | 0.03 | 0.05 |
| J1847.8+3223 | B2 1846+32A | FSRQ | LSP | 0.798 | 24 | 3.30 | 1.67 | 6.85 | 1.23 | >99.0 | 0.17 | 0.06 |
| J1849.4+6706 | S4 1849+67 | FSRQ | LSP | 0.657 | 46 | 6.31 | 4.99 | 19.62 | 1.86 | >99.0 | 0.60 | 0.05 |
| J1911.2−2011 | PKS 1908−201 | FSRQ | LSP | 1.119 | 31 | 4.35 | 3.01 | 13.95 | 1.77 | >99.0 | 0.40 | 0.07 |
| J1923.3−2101 | TXS 1920−211 | FSRQ | LSP | 0.874 | 41 | 6.18 | 3.86 | 23.52 | 2.25 | >99.0 | 0.34 | 0.04 |
| J2000.2+6506 | 1ES 1959+650 | BLLac | HSP | 0.047 | 43 | 2.70 | 1.19 | 5.80 | 2.32 | 94.8 | 0.05 | 0.04 |
| J2009.4−4850 | PKS 2005−489 | BLLac | HSP | 0.071 | 30 | 1.96 | 0.62 | 3.53 | 1.71 | 1.0 | −0.11 | 0.05 |
| J2025.6−0736 | PKS 2023−07 | FSRQ | LSP | 1.388 | 40 | 7.60 | 5.07 | 19.39 | 1.85 | >99.0 | 0.42 | 0.04 |
| J2056.1−4715 | PKS 2052−47 | FSRQ | LSP | 1.491 | 37 | 3.41 | 1.66 | 8.96 | 1.41 | >99.0 | 0.15 | 0.04 |
| J2139.4−4238 | MH 2136-428 | BLLac | ISP | >0.24 | 45 | 2.71 | 1.18 | 5.60 | 1.34 | 98.4 | 0.06 | 0.04 |
| J2143.2+1741 | OX 169 | FSRQ | LSP | 0.213 | 41 | 3.75 | 1.87 | 9.95 | 1.47 | >99.0 | 0.17 | 0.04 |
| J2147.1+0931 | OX 74 | FSRQ | LSP | 1.113 | 45 | 3.32 | 1.66 | 8.93 | 1.18 | >99.0 | 0.17 | 0.04 |
| J2157.5+3125 | B2 2155+31 | FSRQ | LSP | 1.486 | 28 | 2.12 | 0.72 | 4.01 | 1.01 | 8.4 | −0.06 | 0.05 |

TABLE 1—*Continued*

| 0FGL | Other Name | Optical Class | SED Class | $z$ | $N_p$ | $\langle F_{300}\rangle$[a] | $S$[a] | $F_M$[a] | $\sigma_M$[a] | Prob | $\sigma^2_{NXS}$ | err($\sigma^2_{NXS}$) |
|---|---|---|---|---|---|---|---|---|---|---|---|---|
| J2158.8−3014 | PKS 2155−304 | BLLac | HSP | 0.116 | 47 | 7.89 | 2.38 | 14.93 | 1.75 | >99.0 | 0.06 | 0.01 |
| J2202.4+4217 | BL Lac | BLLac | LSP | 0.069 | 42 | 4.26 | 2.31 | 13.90 | 1.90 | >99.0 | 0.22 | 0.04 |
| J2203.2+1731 | OY 101 | FSRQ | LSP | 1.076 | 34 | 2.83 | 1.29 | 5.49 | 1.28 | >99.0 | 0.10 | 0.05 |
| J2207.0−5347 | PKS 2204−54 | FSRQ | LSP | 1.215 | 11 | 2.44 | 2.65 | 10.71 | 1.40 | >99.0 | 1.17 | 0.24 |
| J2229.8−0829 | PKS 2227−08 | FSRQ | LSP | 1.560 | 41 | 3.71 | 1.40 | 7.05 | 1.16 | >99.0 | 0.07 | 0.03 |
| J2232.4+1141 | CTA 102 | FSRQ | LSP | 1.037 | 43 | 3.10 | 1.50 | 9.11 | 1.34 | >99.0 | 0.15 | 0.04 |
| J2254.0+1609 | 3C 454.3 | FSRQ | LSP | 0.859 | 46 | 27.36 | 24.58 | 94.73 | 3.91 | >99.0 | 0.82 | 0.02 |
| J2325.3+3959 | B3 2322+396 | BLLac | LSP | $\cdots$ | 20 | 2.07 | 1.33 | 5.32 | 1.34 | >99.0 | 0.26 | 0.11 |
| J2327.3+0947 | OZ 42 | FSRQ | LSP | 1.843 | 42 | 3.05 | 2.12 | 13.67 | 1.39 | >99.0 | 0.40 | 0.06 |
| J2345.5−1559 | PMN J2345−1555 | FSRQ | LSP | 0.621 | 19 | 2.29 | 0.93 | 4.47 | 1.03 | 69.3 | 0.02 | 0.06 |

[a]Flux ($E > 300$ MeV) units: $10^{-8}$ ph cm$^{-2}$ s$^{-1}$.

## 4. Characterization of temporal variability in weekly light curves

For the first time *Fermi* LAT is enabling the long-term view on high-energy source variability on a uniformly selected sample of gamma-ray sources. In this section, we report the first and quantitative outlook to the 11-month weekly light curves shown in previous section. As mentioned previously, 84 of the LBAS sources have at least a 60% of the 47 weekly bins (i.e. at least 28 weekly bin) with $TS \geq 4$ flux detections (filled points in Figures 1, 2, 3, and 4). This allowed a quantitative time series analysis along the entire light curve (global analyses such as PDS or autocorrelation that are distinct from local analysis as the flare shape analysis reported in Section 6, or wavelet analysis). In particular the Discrete Auto Correlation Function (DACF) and the first-order Structure Function (SF) are suitable methods to provide these first insights on fluctuation modes and characteristic timescales. To reduce contamination in results caused by the low brightness and non-variable sources that provide a white-noise contribution, a sub-sample of 56 brightest and more variable objects is extracted from this list based on variability probability of $\chi^2$ greater than 99% and normalized excess variance $\sigma^2_{NXS} \geq 0.09$ (with exception of Mkn 421, 0FGL J1104.5+3811, and PKS 2155−304, 0FGL J2158.8-3014, taken into account because of their persisting level of flux over the considered period: see Fig. 1 and 4 respectively, and Table 1).

In Figure 12 the observed maximum of the weekly flux variations in subsequent bins is plotted against the redshift (known for 53 of the 56 sources selected). The brightest blazars showing also the most violent variations on weekly timescales, during these first 11 months of *Fermi* survey, are FSRQs PKS 1510−08 (Marscher et al. 2010; Abdo et al. 2010i), PKS 1502+106 (Abdo et al. 2010d), 3C 454.3 (Abdo et al. 2009c), 3C 279 (Abdo et al. 2010g), PKS 0454−234, and ISP BL Lac object 3C 66A. In a few cases

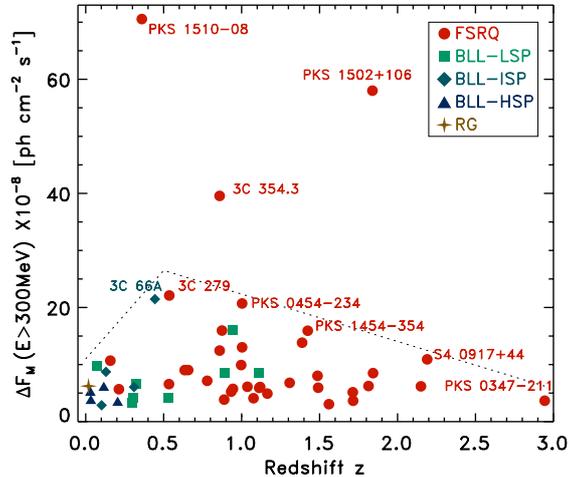

Fig. 12.— Scatter plot of the observed maximum of subsequent weekly flux variations versus the redshift for the 53 brightest and variable sources. Most scattered sources are labeled. Blazars are divided with different symbols and color for the different sub-classes according to table 1.

other BL Lac objects showed rather violent gamma-ray variations, such as AO 0235+164 and BL Lacertae where flux increases approximatively around or above $10^{-7}$ ph cm$^{-2}$ s$^{-1}$ at $E > 300$ MeV. In agreement with results of Section 3 and spectral results reported in (Abdo et al. 2010b), the HSP BL Lac objects are well separated being at the lowest redshift and least variable sources. Apart from the 3 brightest and most variable FSRQs, the other 50 sources appear distributed with decreasing observed maximum gamma-ray variation with increasing redshift as expected by inverse square law. The transition region between the two families is roughly placed between redshift 0.5 and 1. The analysis of the DACF and SF techniques is applied to this same sample of the LBAS list.

The DACF allows to investigate the level of auto-correlation also in discrete data sets (see, e.g. Edelson & Krolik 1988; Hufnagel & Bregman 1992) without any interpolation and any invention of artificial data points. The pairs $[F(t_i), F(t_j)]$ of a discrete data set are first combined in *unbinned* discrete correlations

$$C_{ij}^{(u)} = \frac{(F_i - \langle F \rangle)(F_j - \langle F \rangle)}{\sigma_F^2}, \quad (3)$$

where $\langle F \rangle$ is the average values of the sample and $\sigma_F$ is the standard deviation. Each of these correlations is associated with the pairwise lag $\Delta t_{ij} = t_j - t_i$ and every value represents information about real points. The DACF is obtained by binning the $C_{ij}^{(u)}$ in time for each lag $\Delta t$, and averaging over the number $M$ of pairs whose time lag $\Delta t_{ij}$ is inside $\Delta t$:

$$C(\Delta t) = (1/M) \sum_{ij} C_{ij}^{(u)}. \quad (4)$$

The choice of the bin size for irregular time series is governed by trade-off between the desired accuracy in the mean calculation and the desired resolution in the description of the correlation curve. In this analysis the bin is chosen equal to the sampling, 1 week, because of the limited temporal range and regularity (no gaps) of the light curves.

The SF is equivalent to the power density spectrum (PDS) of the signal calculated in the time domain instead of frequency space, which makes it less subject to sampling problems in presence of very irregular time series, such as windowing and aliasing, (see, e.g. Simonetti, Cordes, & Heeschen 1985; Smith et al. 1993; Lainela & Valtaoja 1993; Paltani et al. 1997). This function represents merely a measure of the mean squared of the flux differences at times $t$ and $t + \Delta t$ of $N$ pairs with the same time separation $\Delta t$, along the whole time series. The first-order SF is defined as

$$\text{SF}^{(1)}(\Delta t) = \frac{1}{N} \sum_{i=1}^{N} \left[ F(t_i) - F(t_i + \Delta t) \right]^2. \quad (5)$$

(where $F_i$ is the discrete signal at time $t$). The general definition involves an ensemble average. This function is a sort of "running" variance of the process that is able to discern the range of timescales that contribute to variations in the time series.

In the DACF and SF analysis of these 56 weekly LBAS light curves, true upper limits ($TS < 1$) are conservatively considered as values close to zero (i.e. $10^{-12}$ phot cm$^{-2}$ s$^{-1}$, well below the 11-month LAT sensitivity) obtaining the twofold goal of have still evenly sampled time series and avoid the bias in results caused by dropping out completely such bins replacing them by gaps. The comparison of results using the blind SF analysis for the evaluation of the power-law index (see below) as test to light curves taking into account upper limits as explained above and replacing them with gaps, attests a very low difference (about 80% of the sources show a difference in the calculated SF slope between $-0.05$ and $0.05$).

Examples of 12 DACF and SF functions applied to these weekly light curves are reported in Figure 13. They show different auto-correlation patterns, different central peak amplitude, different temporal trends and slopes in logarithmic SF representation, pointing out different variability modes and timescales. The time lag $\Delta t_{cross}$ where the DACF value crosses zero for the first time can indicate the maximum correlation scale, while equally spaced and repeated peaks in the function shape can point out characteristic timescales and hints for possible periodicity. Deep drops of the SF value can mean a small variance and provide again possible signature for a characteristic time scale. The ideal SF increases with the lag $\Delta t$ in a log-log representation like in the plots shown in Figure 13. PDSs of blazars' light curves usually show power-law dependence on the signal temporal frequency $f$ in a wide range of frequencies ($P(f) \propto 1/f^\alpha$). In case of sufficient sampling, sufficient total time range and low noise the SF can show a steep linear trend in a certain range of lags in logarithmic scale with index simply related to the PDS power index by $S \propto (\Delta t)^{\alpha-1}$ (Hughes et al. 1992; Lainela & Valtaoja 1993). If a maximum correla-tion timescale is reached in a light curve, the SF is constant for longer lags, and such turnover point between the power law portion and the constant trend can identify another important characteristic time scale. However, it is sometimes difficult to identify and fit this change of slope, especially for weak sources which can have spurious breaks and wiggling patterns in the SF due to statistical errors (for example, Emmanoulopoulos et al. 2010).

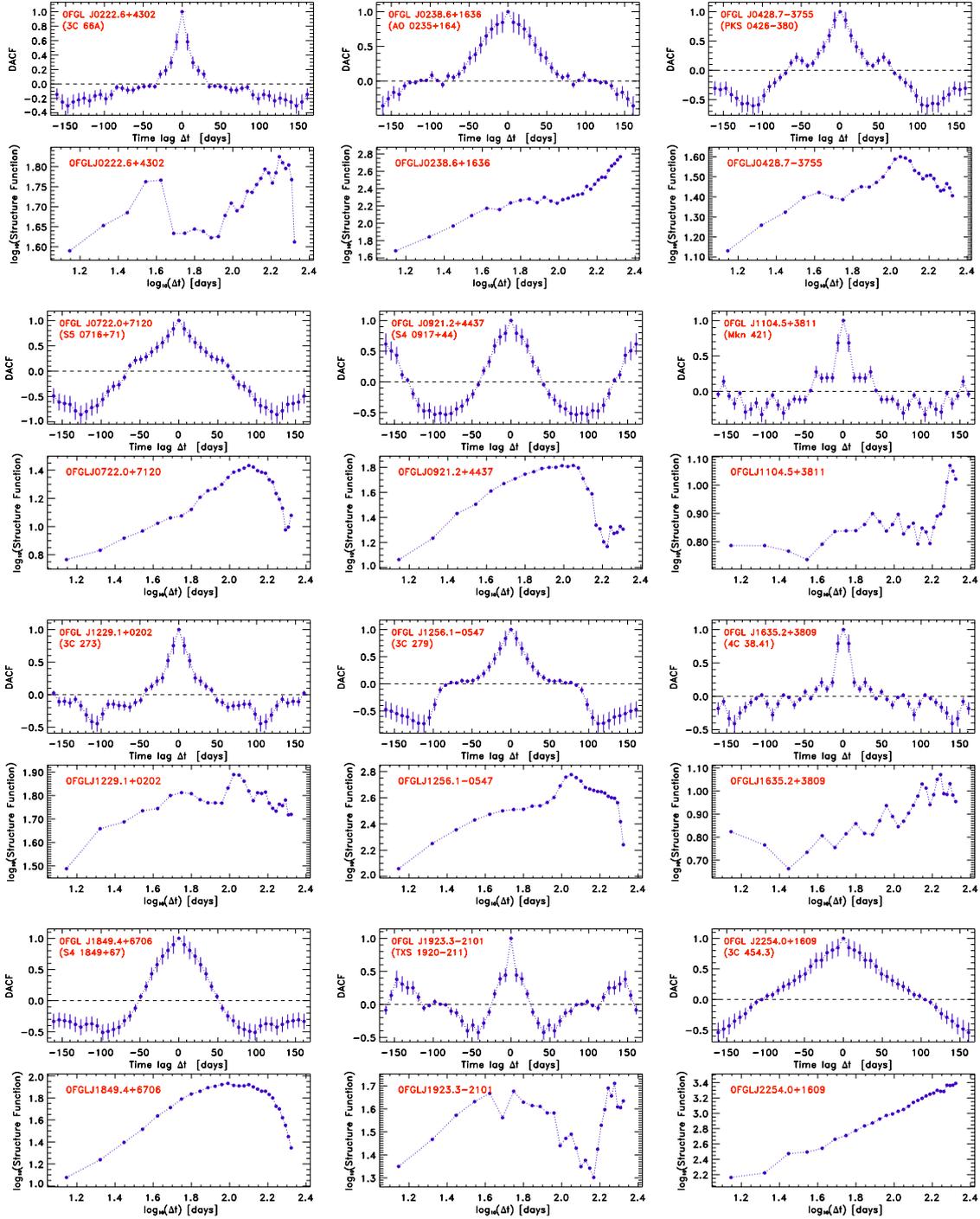

Fig. 13.— Example of DACF and SF functions applied to the weekly light curves of 12 LBAS blazars (from the left to bottom right: 3C 66A, AO 0235+164, PKS 0426−380, S5 0716+71, S4 0917+44, Mkn 421, 3C 273, 3C 279, 4C 38.41, S4 1849+67, TXS 1920+211, and 3C 454.3). They show different auto-correlation patterns, different zero lag peak amplitudes and crossing times, and different temporal power spectral trends and slopes, pointing out more different variability modes.

In Figure 14 we show four distributions of the power indices $\alpha$ evaluated through the SF applied in a blind mode to each of the selected 56 light curve from the minimum lag $\Delta t_{min}$ of 1 week to a maximum lag $\Delta t_{max}$ of 1/3, 1/2, 2/3 and 4/5 of the total time-range $I(= t_{max} - t_{min})$. Most of the $\alpha$ values are distributed between 1.1 and 1.6, meaning a fluctuation mode about halfway between the pure flickering (also known as red-noise, $\alpha = 1$) and the pure shot-noise (also known as brown-noise or Brownian variability, $\alpha = 2$, typically produced by a random walk process). Weaker sources, more affected by error dispersion, cause a whitening of the variability and shift the distributions closer to flickering, as well as the blind application of the SF when the maximum lag adopted is above the function break. For example for the weekly light curve of 3C 279 this power index $\alpha$ estimated by the average on the four blind SF runs is 1.25, whiter than the value found adopting $\Delta t_{max} = (1/3)I$ only (well below the break) where we have a value of about 1.6 (Fig. 14 top panel), in agreement with the value for the 3-day bin light curve found with the direct calculation of PDS analysis (Section 5).

These blind SF results at mid and long-term timescales appear roughly in agreement with the observed long-term optical variability based on some samples (for example optical spectral slope in the range 1.3-1.8 in Heidt & Wagner 1996; Webb & Malkan 2000; Fiorucci et al. 2003; Ciprini et al. 2007), and short-term X-ray variability (e.g. Green et al. 1993; Lawrence & Papadakis 1993; McHardy 2008). Radio light curves have power spectra with slopes around 2 in time-scales from days up to some years (e.g. Hufnagel & Bregman 1992; Hughes et al. 1992; Lainela & Valtaoja 1993; Aller et al. 1999). On the other hand systematic and complete radio/optical studies based on more than one instrument to compare with our gamma-ray variability results are missing.

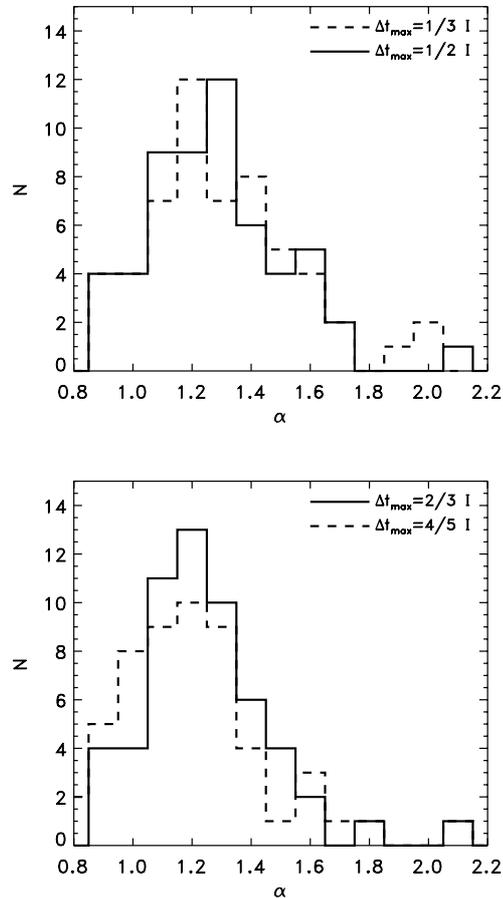

Fig. 14.— Distributions of the PDS power indexes $\alpha$ for the weekly light curves of the 56 most bright and variable LBAS sources, selected as explained in the text. The values are obtained applying the SF considering 4 maximum lags (1/3, 1/2, top panel, and 2/3 and 4/5 bottom panel, of the total time-range $I = t_{max} - t_{min}$). These distributions are peaked for values of the power index between 1.1 and 1.6.

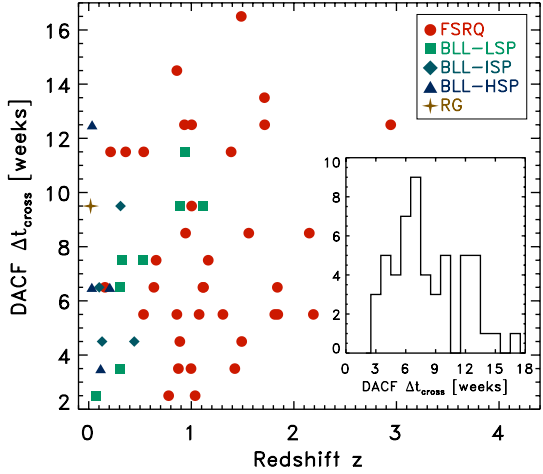
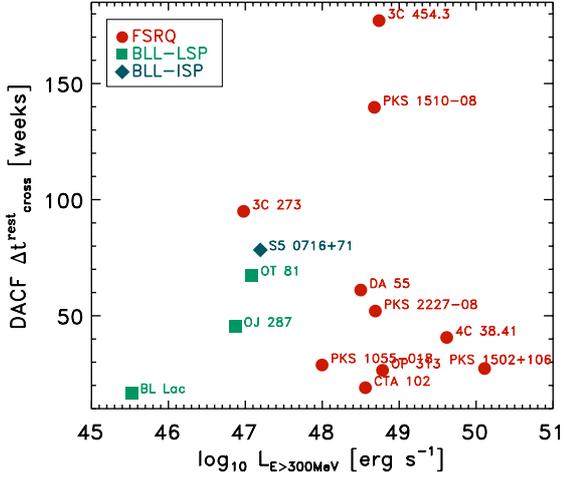
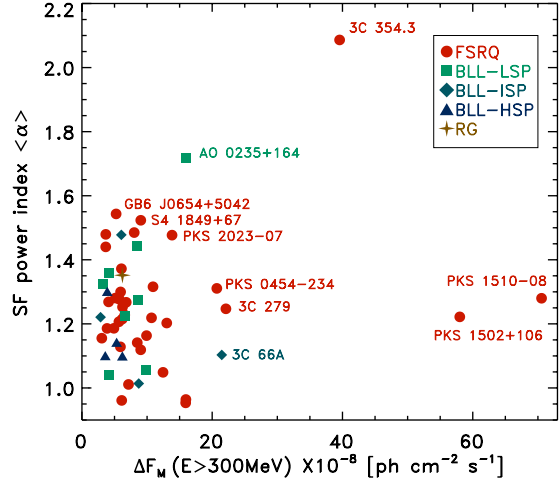

Fig. 15.— *Upper panel*: scatter plot of the DACF crossing times in weeks, versus the redshift for the 53 brightest and variable LBAS sources with a known redshift (or lower limit). The inset panel reports the distribution of values (in weeks) for the same set, pointing out more common $\Delta t_{cross}$ from 4 to 13 weeks and peaked at 7 weeks. *Lower panel*: scatter plot of the DACF crossing times (in weeks) in the rest frame corrected for redshift and relativistic beaming, versus the bolometric absolute gamma-ray luminosity above 300 MeV evaluated taking into account the Doppler and Lorentz factors reported in Savolainen et al. (2010) for 15 of the most bright and variable LBAS sources having factors calculated by MOJAVE observations. All are labeled.

Fig. 16.— Scatter plot of the PDS power index $\alpha$ evaluated in time domain with the SF (averaged among the runs with 4 different time lags) versus the observed maximum of the week-to-week flux variations. Most scattered blazars are labeled.

Remarkable are the cases of 3C 454.3 (0FGL J2254.0+1609 a typical FSRQ) and AO 0235+164 (0FGL J0238.6+1636, LSP BL Lac object) that showed a full Brownian ($\alpha \geq 2$) variability, with a monotonic baseline trend at long timescales, as depicted by the two outliers of Figure 14 and Figure 16 and as shown by the corresponding source light curves, DACF and SF profiles of Figures 1, 4 and 13.

In Figure 15 the DACF crossing times for the 53 brightest and variable sources with known redshift (or lower limit) are plotted against $z$. The distribution is reported as well. The most common $\Delta t_{cross}$ values are from 4 to 13 weeks, pointing out the duration of the autocorrelation and therefore a possible characteristic timescale. The peak bin (9 sources) corresponding to 7 weeks ($\sim$ 49 days) is likely associated with the periodic modulation in efficiency produced by the 55 day precession period of the *Fermi* spacecraft orbit (Abdo et al. 2010h). This is more evident for weakly variable sources such as Mkn 421 and W Com for example, even if intrinsic variability can in principle appear also at these timescales (as could be the case of 3C 273). Characteris-tic timescales can be better searched and quantified using a better sampling as described in the next Section 5. For 15 LBAS sources, that are also in the MOJAVE database, the DACF crossing times are compared with the bolometric intrinsic gamma-ray luminosity above 300 MeV in Figure 15, which is calculated taking into account Doppler and Lorentz factors reported in Savolainen et al. (2010). In particular PKS 1502+106 (z=1.8385) has the record of the most intrinsically violent and gamma-ray luminous outburst shown during these first 11 months of *Fermi* survey (for details see Abdo et al. 2010d). 4C 38.41 (S4 1633+38) and 3C 454.3 (Abdo et al. 2009c) are the other two most powerful gamma-ray blazars in this period.

The values of $\alpha$ averaged over the 4 maximum lags runs of the SF are reported for all the 56 sources in Figure 16 against the maximum of subsequent week-to-week flux variations. In this case a separation between FSRQ and BL Lacs is not evident, but the difference between the variability behavior of full Brownian gamma-ray sources like AO 0235+164 and 3C 454.3 and the variability behavior of other powerful gamma-ray blazars like PKS 1510-08 and PKS 1502+106 is clear. FSRQs like PKS 1510-08 and PKS 1510+106 are characterized by more de-trended flares (departing from a constant baseline level) or by intermittence, while the most apparently bright FSRQs, like 3C 454.3, has clear long term trends and stochastic long term memory (i.e., high-order correlation structure meaning a persistent temporal dependence between observations widely separated in time and low-frequency dominated PDS).

In these weekly light curves no evident sign of periodicity is found, but a more detailed investigation for this aspect will be presented elsewhere using better sampled light curves over only the brightest blazars. In the following two sections a global analysis (PDS) and a local analysis (functional fit of the flare temporal structure) is applied to more densely sampled light curves (3-day and 4-day bins, integrated flux $E > 100$ MeV) extracted only for the brightest 28 sources and 10 sources of the LBAS sample respectively. These light curves, starting from a lower energy threshold because of the high brightness and higher statistics are built as described in Section 2.

## 5. Power Density Spectra of the Brightest Blazars

In lightcurves with binning of a few days, about 15 of the sources are continously, or almost continously, detected throughout the 11 month period. For these 15 sources (9 FSRQs and 6 BL Lacs) we used lightcurves with 3 day binning and for an additional 13 sources (all FSRQs) with slightly lower detection TS we used lightcurves with 4 day binning. All lightcurves were evenly sampled without any data gaps and a Fourier transform routine was used to compute power density spectra (PDS).

The power density is normalized to fractional variance per frequency unit ($rms^2$ $I^{-2}$ $Day^{-1}$) and the PDS points are averaged in logarithmic frequency bins. The white noise level was estimated from the rms of the flux errors and was subtracted for each PDS.

In this section we present resulting PDS for a set of individual sources and also the averaged PDS for the two classes, FSRQs and BL Lacs.

There are a number of effects that can, potentially, distort the PDS of our analysis from the "true" long term variability pattern. This includes stochastic variability within a finite length of observation, systematics in the data and statistical noise. The last effect dominates at high frequencies so for the determination of PDS slopes we use primarily frequencies up to 0.02 $day^{-1}$.

The statistical (measurement) errors in the likelihood based lightcurves were investigated by simulations. These errors were also checked by comparing some lightcurves with corresponding ones obtained by direct aperture photometry, for which Poisson statistics is valid. This showed that the uncertainty in error estimates is not a sig-nificant problem for the brightest sources. For the less bright ones, including all the BL Lacs, this effect does introduce an uncertainty in the estimate of the white noise level in the PDS. The influence of this uncertainty on the PDS slope was estimated by repeating the analysis for a range of possible white noise levels and also by analysis of lightcurves extracted with different time bins (from 1 to 7 days).

Observational and instrument systematics were investigated by analysing pulsar lightcurves extracted from the 11 month data with the same procedure as for the blazars. The most prominent effect is a periodic modulation that is identified with the 55 day precession period of the Fermi spacecraft orbit. This precession is consistent with the addition of the systematic error caused by the variation in effective area due to charged particles during orbital precession. This variation in the LAT effective of area is a known effect that is caused by a change in exposure over the orbital precession period (Abdo et al. 2010h). In the PDS for individual blazars this peak is often hidden by the stochastic variability but does show up when averaging the PDS of a number of sources. The frequency bin at this period was not used when PDS slopes were estimated.

The PDS for some of the brighter sources are shown in Figure 17. The source to source differences are most likely dominated by the stochastic nature of the variability process and there is no significant evidence for a break in any of these cases.

To reduce the stochastic and statistical fluctuations and study the shape of the PDS for FSRQs and BL Lacs as groups we averaged the PDS for each of these two classes using all sources detected with TS > 4500. We do this under the assumption that the differences in PDS shape is small compared to the random fluctuations expected due to the action of the (presumed) underlying stochastic process. The resulting averaged PDS for the 9 FSRQs is shown in Figure 18. The error bars are asymmetric 1 sigma errors for the

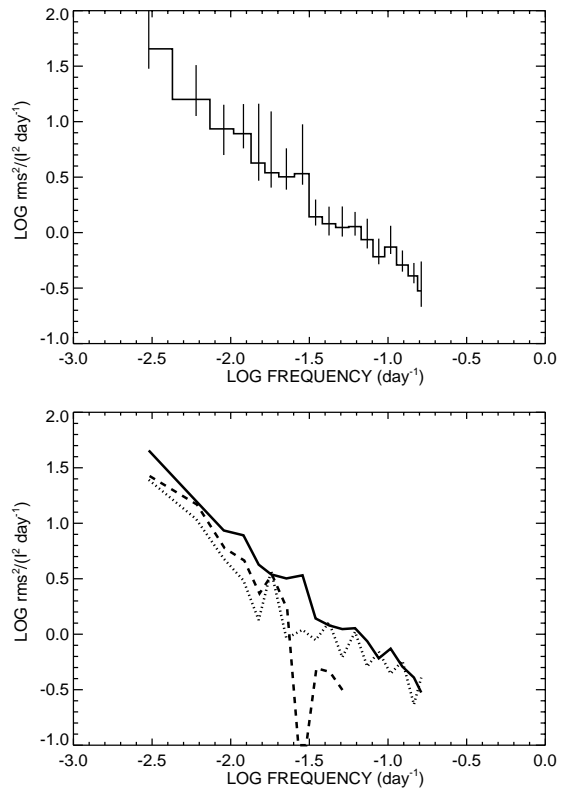

Fig. 18.— Top: Average Power Density spectrum, PDS, for the 9 brightest FSRQs. White noise level based on lightcurve error estimates has been subtracted. The error bars are asymmetric 1 sigma errors of the mean. Our best fit estimate is a PDS slope of 1.4 +/-0.1. Lower: A comparison of the averaged PDS for three sets of sources, the 9 bright FSRQs from the upper plot (solid line), the 6 brightest BL Lac's (dotted line) and 13 additional FSRQs with TS > 1000 (dashed line). Best fit slope for the BL Lac and fainter FSRQs is 1.7 +/-0.3 and 1.5 +/-0.2 respectively

mean over all sources and frequency points averaged in a logarithmic bin. For determination of the PDS slope we focus on the low frequency part, below 0.02 $day^{-1}$, since at higher frequencies the PDS is more sensitive to systematics due to uncertainties in the white noise contribution. For frequencies below 0.017 we obtain a best fit slope of

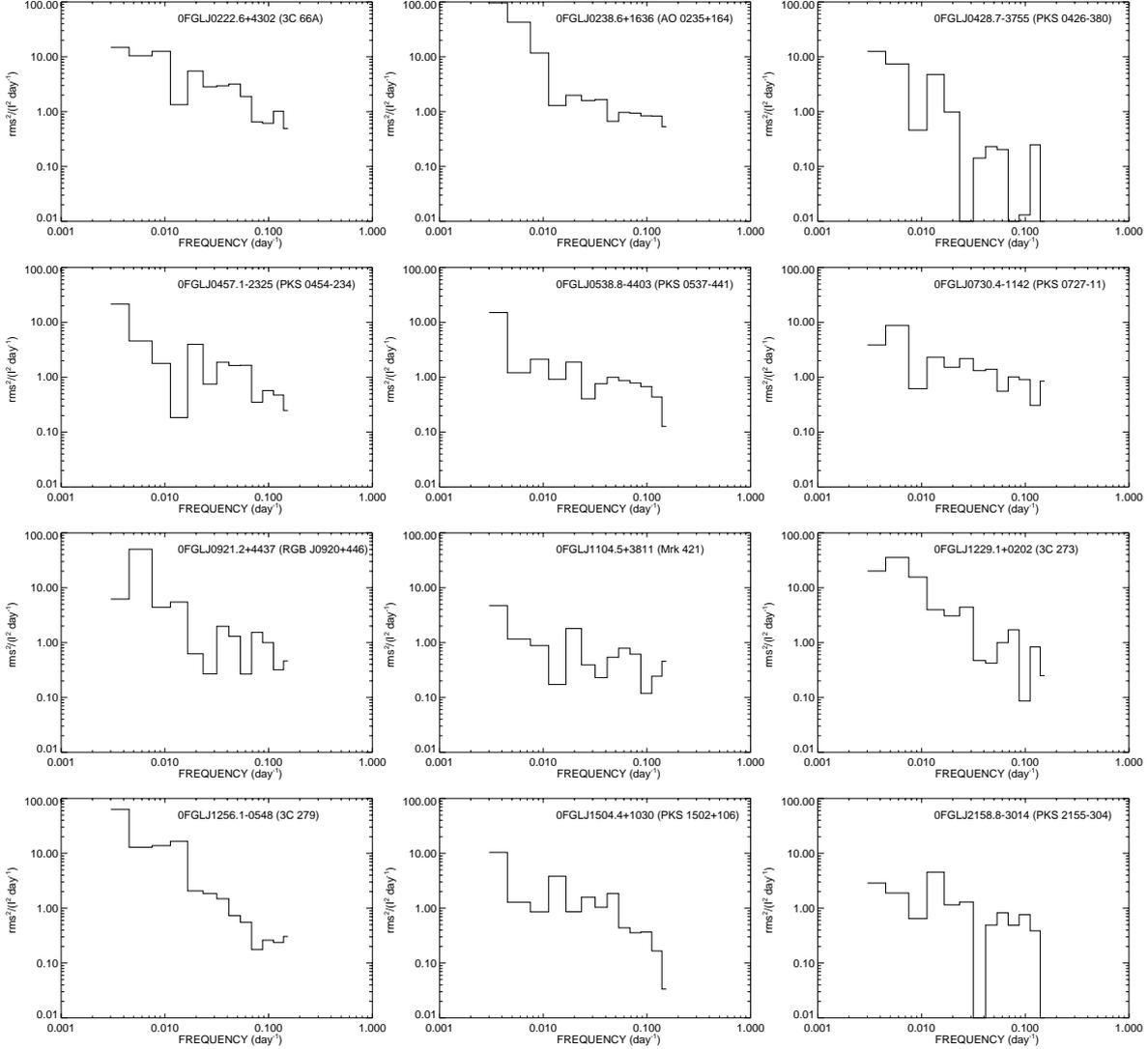

Fig. 17.— Power density spectra computed from 3 day binned lightcurves for some of the brighter sources. The power density is normalized to $rms^2\ I^{-2}\ Day^{-1}$ and the estimated white noise level has been subtracted.

1.4 +/-0.1.

The averaged PDS for the 6 bright BL Lacs is similarly fitted with a power law up to 0.017 $day^{-1}$. This gives a slope of 1.7 +/- 0.3 with white noise based on the lightcurve errors. The sample of sources consists of three LSP's (AO 0235+164, PKS 0426−380 and PKS 0537−441), one ISP (3C 66A) and two HSP's (Mkn 421 and PKS 2155−304). Due to the stochastic nature of the variability and the fact that few sources are considered, it is difficult to draw firm conclusions about the differences between the low and high peaked BL Lacs. An indication of a trend however, is that the three LSP's show stronger variability at longer timescales and therefore dominates the determination of the average, steep slope while the two HSP's both have PDS slopes flatter than 1.0. Further observations are needed to see if this trend can be firmly established.

To increase the data sample and to test if source brightness affects the analysis we selected the remaining FSRQs with TS > 1000 and extracted lightcurves with 4 day binning. Sources where parts of the lightcurve had very large flux errors were not used. This resulting sample consisted of 13 sources for which the PDS was averaged and analysed in the same way as for the brightest sources. For this PDS we obtain best fit slope of 1.5 +/-0.2, in good agreement with the slope for the first sample.

Figure 18 show all three averaged PDS together for comparison. The difference in PDS slope for BL Lacs and FSRQs is of marginal significance but we note that the BL Lac slope is consistent with 2 while this is not the case for the FSRQs. None of the averaged PDS show any significant evidence for the presence of a break although this may still not be excluded for individual sources. From Figure 18 it is also evident that for the present data the fractional variability of the BL Lacs is less than that of the FSRQs, at least up to the 54 day satellite precession peak. The total fractional rms integrated up to 0.017 $day^{-1}$ in the PDS for the 9 FSRQs is 1.35 times that of the 6 BL Lacs. If the ratio is instead estimated by dividing the PDS for the two groups point-by-point (which gives equal weight to each frequency point) we get a value of 1.5. Both estimates were made after subtraction of a white noise level corresponding to the flux error values. If the actual white noise level is larger than this, the ratio between FSRQ and BL Lac fractional variance is most likely larger than our estimate here.

## 6. Temporal Structure of Flares for the Brightest Blazars

The analysis of individual flares is performed using the extracted 3-day time bin flux ($E > 100$ MeV) light curves (except for PKS 1502+106 for which we chose 7-day time bins), as described in Section 2. For this analysis, we selected the light curves of the 10 sources which exhibited high variability with several flares either separated or partially superimposed (see Tables 2 and 3).

We use the following function to reproduce the time profile of a single flare:

$$F(t) = F_c + F_0 \left( e^{\frac{t_0-t}{T_r}} + e^{\frac{t-t_0}{T_d}} \right)^{-1} \quad (6)$$

where $F_c$ represents an assumed constant level underlying the flare, $F_0$ measures the amplitude of the flares, $t_0$ describes approximatively the time of the peak (it corresponds to the actual maximum only for symmetric flares), $T_r$ and $T_d$ measures the rise and decay time. This function is well suited to study both the duration and symmetry of the individual flares. Double exponential forms for the functional fit were used in the past to fit individual blazar flare pulses (Valtaoja et al. 1999). Other and more general functions are used in gamma-ray burst science (see, for example Norris et al. 2000, 2005; Vetere et al. 2006). The time of the maximum of a flare can be easily computed from the first derivative of Equation (6):

$$t_m = t_0 + \frac{T_r T_d}{T_r + T_d} \ln\left(\frac{T_d}{T_r}\right) \quad (7)$$

which is equal to $t_0$ for $T_d = T_r$. A good estimate of the total duration of the flare is:

$$T_{fl} \simeq 2(T_r + T_d) \quad (8)$$

which, for symmetric flares, corresponds to the interval where the flux level is reduced to about 20% of the peak value.

As a first step, we identify the flare to be fitted and detect the time of the peak, which was kept frozen in the fitting procedure unless the flare was clearly superimposed on to a slow trend. We build a function with as many components as the flares' number and perform a fit for each source with the function of Equation 6. To verify the validity of this procedure we analyzed the distribution of the residuals, calculated by subtracting the observed flux from the modeled one and dividing by the flux errors, which should be compatible with a constant level. Figure 19 shows the light curves of six sources with the fit function superimposed, 3C 66A, PKS 0426−380, PKS 0454−234, 3C 273, 3C 279 and 3C 454.3. This procedure was satisfactory for the majority of the flares, but for a few events it did not provide quite good fits. For instance, in the case of the first flare 3C 273 some data points lie above the fitting curve and this discrepancy could be due to events of short duration which were not well sampled.

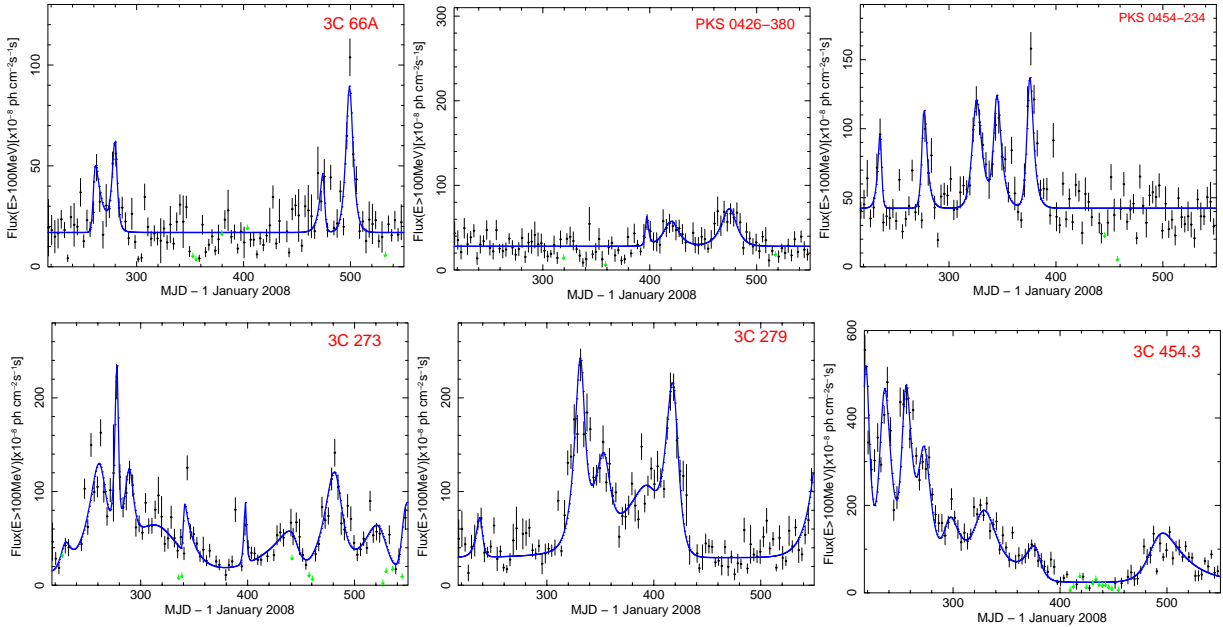

Fig. 19.— Six representative light curves ($E > 100$ MeV) of bright blazars (3C 66A, PKS 0426−380, PKS 0454−234, 3C 273, 3C 279 and 3C 454.3) obtained with 3-day bins. Data points represent detected flux values having a test statistic greater than 9, and the continuous (blue) curve represents the best fit function described in Equation 6.

We defined also the following parameter to describe the symmetry of the flares:

$$\xi = \frac{T_d - T_r}{T_d + T_r} \quad (9)$$

which spans between -1 and 1 for completely right and left asymmetric flares, respectively.

The value of $\xi$ can provide useful indication of the physical evolution of the flare. Those hav-ing a marked asymmetric profile can be explored in terms of a fast injection of accelerated particles and a slower radiative cooling and/or escape from the active region. Symmetric flares, with or without a long standing plateau, can be related to the crossing time of radiation (or particles) through the emission region or can be the result of the superposition of several episodes of short duration. The $\xi$ parameter is used to define three different classes of flares: $i)$ symmetric flares where $-0.3 < \xi < 0.3$, $ii)$ moderately asymmetric flares when $-0.7 < \xi < -0.3$ or $0.3 < \xi < 0.7$ and $iii)$ markedly asymmetric flares when $-1.0 < \xi < -0.7$ or $0.7 < \xi < 1.0$. The parameters are listed in Tables 2 and 3, and their distributions are shown in Figure 20.

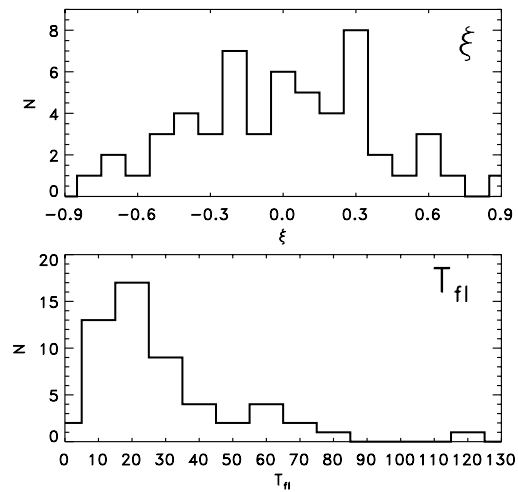

Fig. 20.— Distributions of the flare pulse parameters for the cumulated 10 bright blazars analyzed with this technique. Values of $\xi$ above and $T_{fl}$ below are shown.

Table 2: Summary of the flare structure fit of the brightest blazars using their 3-day bin light curves

| 3C 66A | $\chi_r^2 = 8.0$ | |
|---|---|---|
| flare_time[a] | $\xi$ | $T_{fl}$[b] |
| 260 | 0.73 ± 0.30 | 14.7 ± 11.9 |
| 280 | -0.19 ± 0.19 | 8.3 ± 1.7 |
| 475 | -0.55 ± 0.33 | 8.8 ± 2.5 |
| 495 | 0.02 ± 0.67 | 11.8 ± 7.9 |
| Average | 0.003 ± 0.207 | 10.88 ± 3.6 |
| AO 0235+164 | $\chi_r^2 = 11.4$ | |
| flare_time[a] | $\xi$ | $T_{fl}$[b] |
| 230 | 0.16 ± 0.07 | 21.7 ± 1.5 |
| 251 | 0.61 ± 0.06 | 34.4 ± 1.5 |
| 291 | -0.49 ± 0.11 | 24.0 ± 1.9 |
| 301 | 0.63 ± 0.43 | 21.1 ± 7.9 |
| 380 | 0.04 ± 0.13 | 63.9 ± 8.3 |
| 401 | 0.36 ± 0.09 | 6.3 ± 0.5 |
| 425 | 0.55 ± 0.18 | 47.3 ± 7.2 |
| 460 | -0.04 ± 0.38 | 19.1 ± 7.4 |
| 485 | -0.17 ± 0.08 | 5.4 ± 0.4 |
| Average | 0.18 ± 0.07 | 27.02 ± 1.74 |
| PKS 0426−380 | $\chi_r^2 = 2.2$ | |
| flare_time[a] | $\xi$ | $T_{fl}$[b] |
| 395 | 0.10 ± 0.07 | 4.4 ± 0.3 |
| 420 | 0.01 ± 0.30 | 24.2 ± 7.4 |
| 475 | -0.11 ± 0.07 | 25.2 ± 1.8 |
| Average | 0.0003 ± 0.1068 | 17.97 ± 2.52 |
| PKS 0454−23 | $\chi_r^2 = 5.0$ | |
| flare_time[a] | $\xi$ | $T_{fl}$[b] |
| 236 | -0.39 ± 0.10 | 6.6 ± 0.5 |
| 276 | 0.33 ± 0.24 | 9.0 ± 2.2 |
| 325 | 0.21 ± 0.08 | 16.0 ± 1.2 |
| 344 | 0.30 ± 0.63 | 13.0 ± 7.9 |
| 375 | 0.25 ± 0.78 | 10.9 ± 8.2 |
| Average | 0.14 ± 0.21 | 11.07 ± 2.34 |
| S4 0917+44 | $\chi_r^2 = 9.4$ | |
| flare_time[a] | $\xi$ | $T_{fl}$[b] |
| 364 | 0.0001 ± 0.1650 | 120.1 ± 19.8 |
| 520 | 0.0001 ± 0.0707 | 80.1 ± 5.6 |
| Average | 0.0001 ± 0.0898 | 100.10 ± 10.29 |

[a] day of the flare peak (DoY 2008 unit)

[b] fraction of days

Table 3: Continuation of Table 2.

| 3C 273 | $\chi_r^2 = 3.7$ | |
|---|---|---|
| flare_time[a] | $\xi$ | $T_{fl}$[b] |
| 229 | 0.13 ± 0.07 | 15.8 ± 1.1 |
| 263 | -0.32 ± 0.03 | 30.2 ± 1.3 |
| 278 | -0.24 ± 0.08 | 6.1 ± 0.4 |
| 290 | -0.22 ± 0.17 | 15.5 ± 2.2 |
| 340 | 0.88 ± 0.04 | 17.0 ± 0.4 |
| 398 | -0.11 ± 0.07 | 4.5 ± 0.3 |
| 445 | -0.76 ± 0.04 | 68.0 ± 1.5 |
| 483 | -0.31 ± 0.09 | 30.7 ± 2.3 |
| 525 | -0.46 ± 0.11 | 45.7 ± 3.6 |
| Average | -0.16 ± 0.03 | 25.93 ± 0.89 |
| 3C 279 | $\chi_r^2 = 3.7$ | |
| flare_time[a] | $\xi$ | $T_{fl}$[b] |
| 238 | -0.33 ± 0.24 | 11.0 ± 2.3 |
| 332 | -0.22 ± 0.08 | 17.0 ± 1.2 |
| 355 | -0.41 ± 0.03 | 28.4 ± 1.3 |
| 398 | -0.67 ± 0.05 | 71.9 ± 4.2 |
| 419 | 0.29 ± 0.08 | 22.4 ± 1.6 |
| Average | -0.27 ± 0.05 | 30.11 ± 1.07 |
| PKS 1502+106 | $\chi_r^2 = 4.4$ | |
| flare_time[a] | $\xi$ | $T_{fl}$[b] |
| 242 | -0.20 ± 0.10 | 40.7 ± 3.8 |
| 305 | -0.71 ± 0.12 | 41.6 ± 2.9 |
| 336 | 0.11 ± 0.11 | 31.6 ± 3.2 |
| 365 | 0.13 ± 0.14 | 21.3 ± 3.4 |
| 405 | 0.28 ± 0.09 | 55.6 ± 4.1 |
| 485 | -0.18 ± 0.08 | 57.2 ± 4.1 |
| 525 | -0.05 ± 0.07 | 36.8 ± 2.6 |
| Average | -0.09 ± 0.04 | 40.67 ± 1.31 |
| PKS 1510−08 | $\chi_r^2 = 17.8$ | |
| flare_time[a] | $\xi$ | $T_{fl}$[b] |
| 260 | -0.52 ± 0.12 | 25.1 ± 2.0 |
| 381 | -0.39 ± 0.10 | 19.7 ± 1.5 |
| 445 | 0.07 ± 0.01 | 24.2 ± 0.3 |
| 480 | 0.25 ± 0.08 | 11.1 ± 0.8 |
| Average | -0.15 ± 0.04 | 20.00 ± 0.66 |
| 3C 454.3 | $\chi_r^2 = 7.3$ | |
| flare_time[a] | $\xi$ | $T_{fl}$[b] |
| 235 | 0.24 ± 0.16 | 20.3 ± 3.8 |
| 255 | 0.29 ± 0.08 | 18.0 ± 1.3 |
| 272 | 0.24 ± 0.36 | 22.2 ± 7.9 |
| 295 | 0.44 ± 0.28 | 32.2 ± 8.2 |
| 327 | 0.25 ± 0.13 | 41.8 ± 4.4 |
| 378 | -0.42 ± 0.10 | 28.5 ± 2.2 |
| 490 | 0.48 ± 0.11 | 59.4 ± 4.7 |
| Average | 0.22 ± 0.07 | 31.79 ± 1.98 |

[a] day of the flare peak (DoY 2008 unit)

[b] fraction of days

We also calculated the weighted mean of these parameters to study the general properties of the time profiles of gamma-ray flares. We obtain $\langle\xi\rangle = -0.084 \pm 0.009$ and $\langle T_{fl}\rangle = 11.87 \pm 0.12$. Looking at the results of the fitting procedure and the weighted means we can see that the list of brighter sources shows two different types of temporal profiles: the sources with a stable baseline with a sporadic flaring activity and the sources with a strong activity with complex and structured features. Based on our analysis we can put 3C 66A, PKS 0426−380, S4 0917+44, PKS 0454−234 in the first class of objects and the remaining 3C 279, 3C 273, 3C 454.3, PKS 1502+106, AO 0235+164 and PKS 1510−08 in the second one, while no evidence of very asymmetric profiles is found. In Figure 19 we report cases of both classes to show the different time profiles. Note that for the majority of events the uncertainties on $\xi$ are small, however, for a few flares of 3C 66A, AO 0235+164 and PKS 0454−23 the resulting asymmetries are not safely estimated. In fact, despite their large values the occurrence of symmetric of moderately asymmetric profiles cannot be excluded within 1 standard deviation.

We found only four markedly asymmetric flares: for 3C 66A (DoY 2008 260 $\xi = 0.73 \pm 0.30$), 3C 273 (DoY 2008 340 and 445, $\xi = 0.88 \pm 0.04$ and $\xi = -0.76 \pm 0.04$, respectively) and PKS 1502+106 (DoY 2008 305, $\xi = -0.71 \pm 0.12$), where two of them have rise times longer that the decays. In the case of 3C 66A the flare was rather short and the resulting uncertainty on $\xi$ is large, therefore no firm conclusion on its shape can be established. The two flares of 3C 273 clearly exhibit different profiles. Note that the highest point of flare at epoch 340 is well above the fitting curve implying the possibility of an even higher value of $\xi$, whereas the subsequent and much longer flare (DoY 2008 445), which has a very well established negative asymmetry, may be due to confusion because of the partial superposition of low amplitude and short events, not individually detectable. 3C 273 exhibited also a couple of exceptional flares in September 2009 (Abdo et al. 2010f), in which it reached a very high level, and the light curves were very finely sampled. In both episodes rise times were shorter than the subsequent decays. Similarly, PKS 1502+106 exhibited a markedly asymmetric outburst in August 2008, resolved with a daily binning (Abdo et al. 2010d).

## 7. Summary and Conclusions

Gamma-ray light curves (Figures 1-5) and variability properties of the 106 LBAS blazars (0FGL list, Abdo et al. 2009a,b), collected during the first 11 months of the all-sky survey by *Fermi* LAT are presented. This represents a first systematic study of gamma-ray variability over a consistent set of homogeneously observed blazars.

The light curves of 84 of these sources have at least $60\%$ of the 47 weekly bins with flux detection of $TS > 4$ ($\gtrsim 2\sigma$), and 56 have also a significant excess variance (Table 1). The low gamma-ray brightness states interposed among the flares are studied as well for the first time, and high flux states do not exceed 1/4 of the total light curve range (most sources being active in periods shorter than 5% of the total light curve duration). FSRQs and LSP/ISP BL Lac objects showed largest variations, as expected, with the high energy SED component peaked at MeV and GeV bands. HSP BL Lac object show lower variability (with exception of ON 325), and their emission is persistent, easily detected in all the weeks of the considered period (Section 3).

In these first 11 months of *Fermi* mission PKS 1510−08, PKS 1502+106, 3C 454.3, 3C 279, PKS 0454−234 (all FSRQs) are observed to be the most bright and violently variable gamma-ray blazars. In a few cases this was true also for BL Lac objects (3C 66A and AO 0235+164 for example). In particular PKS 1502+106 (OR 103), 4C 38.41 (S4 1633+38) and 3C 454.3 were also the most intrinsically gamma-ray powerful blazars in these months. The other sources appear distributed with decreasing observed maximum subsequent variations with increasing redshift. Different auto-correlation patterns, central lag peak amplitudes, zero crossing times, different temporal trends and power-law indices are shown by the DACF and SF, pointing out different timescales and variability modes (more flicker-dominated or Brownian-dominated). The weekly PDSs evaluated using in blind mode the SF point out a $1/f^\alpha$ trend with values mostly distributed between 1.1 and 1.6. Light curves of AO 0235+164 and 3C 454.3 are observed to be fullly Brownian (i.e. with the steepest PDS slopes, $\alpha \geq 2$) with longer emission cycles and sustained flares, that could identify more massive central black holes. Other powerful sources such as PKS 1510-08 and PKS 1510+106 show variability behavior half-way between the two clases above (with $\alpha \sim 1.3$) showing intermittence and de-trended complex superposed flares respectively. The DACF crossing lag times are found mostly distributed between 4 and 13 weeks with a peak at 7 weeks.

The mean variability properties for the brighter sources are studied in more detail by calculating an average PDS for each of the two main blazar types, FSRQs (9 sources) and BL Lacs (6 sources). For both types the average PDS is described by a power law without any evidence for a break in the frequency range where our sensitivity is best (0.003 to at least 0.017 $day^{-1}$). The power law index for the averaged PDS was estimated to $1.4 \pm 0.1$ and $1.7 \pm 0.3$ for the FSRQs and BL Lacs, respectively. The BL Lac sources show a large spread in PDS slopes with an indication of trend such that the PDS is steeper for LSPs than for the HSPs. Further observations are needed to establish this trend, but we note that in the present data the two brighter HSP's (Mkn 421 and PKS 2155−304) have PDS slopes of order 1 or flatter. For Mkn 421 we can compare this with the corresponding result for soft X-rays. Analyzing the RXTE ASM X-ray light curve for this source we obtain a well defined power law index of $1.04 \pm 0.05$. Besides Mkn 421 the best available long-term X-ray light curve is that of the FSRQ 3C 279. For this source Chatterjee et al. (2008) found a slope of $2.3 \pm 0.3$ for the X-ray band and $1.7 \pm 0.3$ for the optical. In this case our result in the gamma-ray band, both the average for FSRQs and for 3C 279 itself ($1.6 \pm 0.2$), is closer to the PDS slope in optical than in X-rays. More generally the gamma-ray PDS of bright *Fermi* LAT sources have slopes similar to those obtained from long-term optical and radio light curves (Section 4). For the X-ray band the situation is less clear since only a few sources have good enough long-term light curve to allow a comparison.

The power density excess (above the noise level) in the 0.003 to 0.017 $day^{-1}$ range was found to correspond to a mean rms fractional variability (rms/$I^2$) of 0.50 for the 9 bright FSRQs and 0.37 for the 6 brightest BL Lac's. These results imply that in the LAT energy range and presently accessible time scales the FSRQs exhibit a larger relative variability than the BL Lac's.

Gamma-ray variability observed in these LBAS blazars can be described both as essentially steady sources with perturbations or as a series of discrete, though possibly overlapping flares produced for example by traveling shock fronts. The emission could be produced in multiple regions forming a inherently inhomogeneous blazar zone or in an essentially homogenous region where all particles are accelerated, depending by the particular source considered.

Random walk processes producing such PDS variability slopes, like instabilities and turbulence in the accretion flow through the disc or in the jet, can cause the intermittent behavior observed in several of these *Fermi* LAT light curves. These are stochastic processes, mostly characterized by the presence of a large number of weakly correlated elements which appear at random, live only a short time and decay. Steep PDS slopes means more Brownian-dominated regimes characterized more by long memory and self-similarity. Large flares likely arise from the sudden acceleration of relativistic electrons, related to bulk injections of new particles and/or strong internal shocks (Mastichiadis & Kirk 1997; Spada et al. 2001; Böttcher & Dermer 2010). These type of PDS could be related to mass accretion avalanches providing shot pulses: larger (and longer) shots contribute to the low-frequency part of the PDS, while smaller and shorter shots determine the power-law decline at high frequencies. In this case variability would be explained as disturbances or inhomogeneities in the accretion process, opposite to intermittence that can be evidence of dissipation in the jet and described by turbulence-driven processes. Furthermore well identified GeV recurrent characteristic timescales, pointed out by breaks in the PDS, can be related linearly to the mass of the central supermassive black hole (Markowitz 2006; Dermer 2007; Wold et al. 2007; McHardy 2008), as happens for X-ray variability timescales in Seyfert galaxies, but more detailed analysis with improved sampling is needed to shed light on this question.

Finally the local analysis of flare temporal shapes for the brightest sources revealed and confirmed in a quantitative way different temporal profiles: stable baseline with sporadic flaring activity or strong activity with complex and structured temporal features. The average durations of the fitted flares varied from about 10 days up to 100 days in the case of S4 0917+44. In other very bright flares, times scales as short as a fraction of day have been observed (3C 273, PKS 1510−089) and in some cases the light curves were structured in series of shorter peaks. The low mean asymmetry of the events analyzed in Section 6 can be then explained by the superposition of series of peaks, even if the light curves analyzed are already resolved with a short, 3-day, sampling. A marked asymmetric profile can be explained in terms of a fast injection of accelerated particles and a slower radiative cooling and/or escape from the active region, and could be considered cooling-dominated flares. The fast rise and slower decay can be evidence for a dominant contribution by Comptonization of photons produced outside the jet (Sikora et al. 2001). Gamma-ray flares produced by short-lasting energetic electron injections and at larger jet openig angles are predicted to be more asymmetric showing much faster increase than decay, the latter determined by the light travel time effects. On the other hand symmetric flares, with or without a long standing plateau, can be related to the crossing time of radiation (or particles) through the emission region, dominated by geometry and spatial scales (Takahashi et al. 2000; Tanihata et al. 2001). Flares observed at or above the peak energy reflect the scale of the source along the line of sight and are symmetric for cylindrical geometry of the active regions (Eldar & Levinson 2000; Sokolov et al. 2004). The result of the superposition and blend of several episodes of short duration could also provide symmetric flare shapes (Valtaoja et al. 1999).

The presentation of gamma-ray light curves of a consistent set of blazars observed in homogeneous conditions by *Fermi* LAT over almost one year, and our systematic variability characterization showed properties in some way similar to the radio-band and optical variability. Variation amplitudes, flare durations, PDS slope values, preliminary hints for typical timescales, and mor-phology of the flares can be used to support the identification of the correct source class for newly discovered unidentified sources. Basically LAT gamma-ray blazar are displaying two "flavors" of variability: rather constant baseline with sporadic flaring activity showing also intermittence and characterized by more flat PDS slopes resembling the red-noise, flickering, fluctuations, and a few sources strong activity with complex and structured time profiles characterized by the long memory and steeper PDS slopes of random-walk processes. Finally, our results can also serve as preparatory study for more detailed analysis and modeling that are possible with the brightest and most variable sources through a better sampling and time resolution.

## 8. Acknowledgments

The *Fermi* LAT Collaboration acknowledges generous ongoing support from a number of agencies and institutes that have supported both the development and the operation of the LAT as well as scientific data analysis. These include the National Aeronautics and Space Administration and the Department of Energy in the United States, the Commissariat à l'Energie Atomique and the Centre National de la Recherche Scientifique / Institut National de Physique Nucléaire et de Physique des Particules in France, the Agenzia Spaziale Italiana and the Istituto Nazionale di Fisica Nucleare in Italy, the Ministry of Education, Culture, Sports, Science and Technology (MEXT), High Energy Accelerator Research Organization (KEK) and Japan Aerospace Exploration Agency (JAXA) in Japan, and the K. A. Wallenberg Foundation, the Swedish Research Council and the Swedish National Space Board in Sweden.

Additional support for science analysis during the operations phase is gratefully acknowledged from the Istituto Nazionale di Astrofisica in Italy and the Centre National d'Etudes Spatiales in France.

*Facilities: Fermi* LAT.